%% file: main.tex
\documentclass[compsoc,conference,a4paper,10pt,times]{IEEEtran}

\IEEEoverridecommandlockouts

\usepackage{cite}
\usepackage{amsmath,amssymb,amsfonts}
\usepackage{algorithmic}
\usepackage{graphicx}
\usepackage{textcomp}
\usepackage{bmpsize}
\usepackage{xcolor}
\usepackage{lipsum}
\usepackage[colorlinks=true,urlcolor=black]{hyperref}
\usepackage[braket, qm]{qcircuit}
\usepackage{filecontents}
\usepackage{algorithm}
\usepackage{caption}
\usepackage{subcaption}
\usepackage{adjustbox}
\usepackage{setspace}
\usepackage{tabularx}
\usepackage{balance}
\usepackage{pifont}
\usepackage{bbding}
\usepackage{multicol,multirow}
\usepackage{color}
\usepackage{soul}

\newcommand{\email}[1]{\href{mailto:#1}{#1}}
\newcommand{\nsf}[1]{\href{https://www.nsf.gov/awardsearch/showAward?AWD_ID=#1}{#1}}

\begin{document}

%%
%% Title
%\title{Securing HHL Quantum Algorithm against Attacks During Execution on Quantum Computers}
\title{Securing HHL Quantum Algorithm\\ against Quantum Computer Attacks}

%%
%% Authors

\author{\IEEEauthorblockN{Yizhuo Tan}
\IEEEauthorblockA{
\textit{Yale University}\\
New Haven, US \\
\email{yizhuo.tan@yale.edu}}
\and
\IEEEauthorblockN{Hrvoje Kukina}
\IEEEauthorblockA{
\textit{TU Wien}\\
Vienna, Austria \\
\email{hrvoje.kukina@student.tuwien.ac.at}}
\and
\IEEEauthorblockN{Jakub Szefer}
\IEEEauthorblockA{
\textit{Northwestern University}\\
Evanston, USA \\
jakub.szefer@northwestern.edu}
}

\maketitle

\begin{abstract}
\input{sections/abstract}
\end{abstract}

%%
%% Main paper sections
\input{sections/introduction}

\input{sections/background}

\input{sections/threat_model}

\input{sections/set_up}

\input{sections/attack}

\input{sections/defense}

\input{sections/simulation}

\input{sections/hardware_attack}

\input{sections/hardware_defense}

\input{sections/upgraded_hhl}

\input{sections/related_work}

\input{sections/conclusion_future_work}

\input{sections/acknowledgements}

% \begin{acks}
% \end{acks}

%\clearpage
\balance
\bibliographystyle{IEEEtranS}
\bibliography{main}

% \appendix

\end{document}

%% file: sections/abstract.tex
As the quantum research community expands and new quantum algorithms are created and implemented, it is essential to consider the security implications and potential threats that could lead to the compromise the information processed by them. This work focuses on securing the HHL quantum algorithm against attacks while it executes on a quantum computer. Specifically, two types of potential attacks could be deployed on a cloud-based quantum computer by an attacker circuit attempting to interfere with the victim HHL circuit: the Improper Initialization Attack (IIA) and the Higher Energy Attack (HEA). To protect the HHL algorithm from IIA and HEA, this work proposes first-of-a-kind defense strategies against these attacks on the HHL quantum algorithm. Next, this work demonstrates an implementation of a new quantum circuit for the HHL quantum algorithm that incorporates these defenses. The redesigned quantum circuit is necessary to successfully apply and realize all proposed defense strategies. Finally, this work illustrates how these defense strategies function in practice in the redesigned circuit, specifically how they can protect the HHL quantum algorithm from both IIA and HEA across multiple qubits involving all three types of qubits used in the HHL algorithms: ancilla, clock, and b. The defense requires minimal modification to the circuit, and has only a very small effect on the fidelity of the circuits. The circuits have been tested and validated in both simulation, and also on real IBM quantum computer hardware. The work further analyzes how the modified HHL circuit with the defenses is affected by noise during quantum computation. This work in the end demonstrates that it is practical to add protections to quantum circuits so that they not only perform correct computation, but also self-detect if an attack has occured during the execution.

%% file: sections/introduction.tex
\section{Introduction}
\label{sec:introduction}

In last few years, there has been a significant surge in the development of quantum computers and availability of  of quantum processing units (QPUs)~\cite{Preskill2018quantumcomputingin}, which can be easily accessed online. Numerous technologies form basis of different types of QPUs, including: superconducting qubits~\cite{ibmquantum}, trapped ions~\cite{quantinuum}, neutral atoms~\cite{quera}, silicon spin qubits~\cite{neyens2024probing}, photons~\cite{zhong2020quantum}, and diamond NV centers~\cite{QTI}. Quantum computers based on these technologies are mostly so-called Noisy Intermediate-Scale Quantum (NISQ) computers, which do not have error correction yet. However, there are already algorithms that researchers are developing which can run on these quantum computers.

Among the promising algorithms for use with NISQ quantum computers is the HHL~\cite{harrow2009quantum} quantum algorithm (named after its authors, Harrow, Hassidim, and Lloyd), particularly its upgraded versions that feature improved computational complexity~\cite{ambainis2010variabletimeamplitudeamplification},~\cite{Childs_2017},~\cite{Wossnig_2018}, and various new hybrid instances~\cite{Cao_2012},~\cite{Lee_2019},~\cite{Zhang2022},~\cite{yalovetzky2024solvinglinearsystemsquantum},~\cite{https://doi.org/10.1002/andp.202200082}. It is designed for generating solutions to a set of linear equations. The HHL algorithm serves as a quantum counterpart to classical linear equation solvers such as Gaussian elimination and conjugate gradient method~\cite{Hestenes1952-rl}. The HHL algorithm can have many applications, including in quantum machine learning~\cite{schuld2015introduction, biamonte2017quantum, DUAN2020126595, shao2018reconsiderhhlalgorithmrelated}.

With advent of cloud-based quantum computers, the algorithms such as HHL, will not be run on stand-alone machines. Rather, the Quantum Processing Unit (QPU) hardware in a cloud-based setting will be shared among different users and algorithms. Already today many cloud providers offer access to cloud-based QPUs, such as with IBM Quantum~\cite{ibmquantum}. The QPU hardware can be time-shared where circuits, or shots of circuits, may be interleaved at different granularities. In future, multi-tenant quantum computers, which are already proposed in research~\cite{d2023distributed, 9840181, upadhyay2023stealthyswapsadversarialswap}, maybe deployed and further extend the amount of sharing among different, possibly distrusting, users. Consequently, security threats in the temporal and spatial sharing scenarios of QPUs need to be explored.

\begin{figure*}[t]
\center{\includegraphics[width=\textwidth]{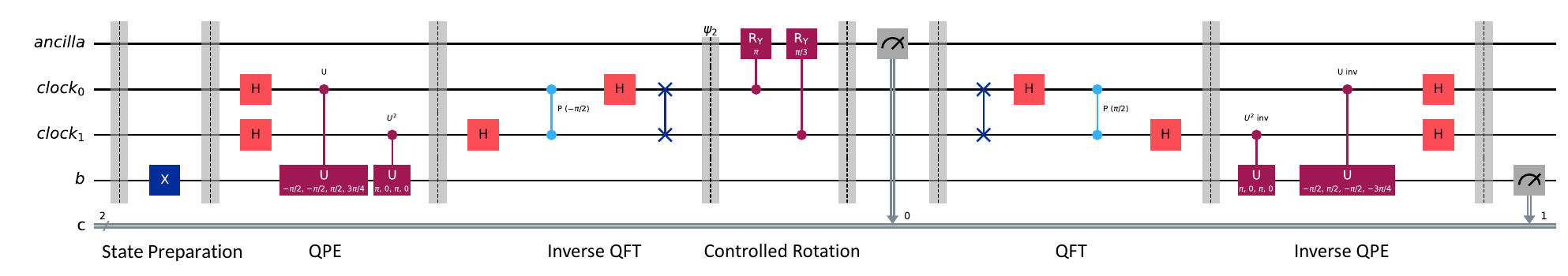}}
\caption{Quantum circuit for the HHL algorithm, the vertical barriers are used to separate the different phases of the algorithm; the phases are labeled at the bottom of the circuit.}
\label{fig:hhl}
\end{figure*}

In a scenario where the adversary and victim may share the same quantum computer, this work explores two types of attacks, their impact on the important HHL algorithm, and potential defense strategies. The first attack is what is called Improper Initialization Attack (IIA), involves malicious setting of the initial state of a qubit to $\ket{1}$, while it is expected that qubit should be set to $\ket{0}$. The second attack is the Higher Energy Attack (HEA)~\cite{xu2023securing}, which involves setting the initial state of a qubit to a higher energy states such as $\ket{2}$ or $\ket{3}$, instead of the expected $\ket{0}$. In both cases, the incorrect setting of the qubit causes, as we demonstrate, incorrect results. Recent work has shown that even changing just one qubit in the victim algorithm is sufficient to launch the attack~\cite{xu2023securing}. The IIA could be launched when attacker is able to influence initial state of the qubit, for example through cross-talk; or the IIA can be part of software supply-chain attack where attacker is able to manipulate user's circuit without user's knowledge. The HEA is based on prior work that has demonstrated that setting a qubit to higher energy causes quantum gates applied to that qubit to not work correctly. An attacker may set a qubit to a higher energy state, and even if qubits are reset between attacker and victim circuit shots, the higher energy state propagates and causes victim's circuit to not work correctly. We perform in-depth evaluation of effect of the attacks on different qubits within HHL algorithm: ancilla qubit, clock qubit, and b qubit. In all cases, malicious output can be achieved even when only one qubit is attacked.

The attacks demonstrate that the HHL algorithm can be manipulated in various ways. Therefore, this work proposes new defense strategies against IIA and HEA. To achieve this, a new design for the HHL quantum circuit has been developed. New quantum registers were added to the original HHL circuit, along with additional quantum gates for the defense of the ancilla, clock, and b qubits. All defense strategies have been tested against real attacks, and the results show that the proposed strategies were able to neutralize the attacks on each qubit, preventing the attacker from disrupting the information and causing damage.

\subsection{Contributions}

The contributions of this work to the field of quantum computer security are as follows.

\begin{enumerate}
  \item Analysis of two types of potential attacks that could be deployed on a cloud-based quantum computers against the HHL algorithm.
  \item Design of defenses for the HHL quantum algorithm against these two types of attacks: IIA and HEA.
  \item Evaluation of attacks and defenses on both simulators and real quantum computers.
\end{enumerate}

%% file: sections/background.tex
\section{Background}\label{sec:background}

This section presents background on the HHL algorithm. It futher gives background on supply chain attacks, which are one means of launching the IIA, and it also overviews higher energy states, which are key to the HEA.

\subsection{HHL Algorithm}

HHL is a quantum algorithm named after its authors, Harrow, Hassidim, and Lloyd, designed for quantum mechanical solutions to linear equations. It may also be referred to as a quantum matrix inversion algorithm. The HHL algorithm serves as a
quantum counterpart to classical linear equation solvers such as Gaussian elimination and the conjugate gradient algorithm~\cite{Hestenes1952-rl}. Its standout feature lies in its capacity to solve $N$ linear equations with $N$ variables (under certain conditions and
limitations) in a time span proportional to $O(log N)$, in contrast to the classical $O(N)$,
resulting in an exponential speedup. The HHL algorithm comprises three main components: quantum phase estimation, eigenvalue inversion rotation, and inverse
quantum phase estimation. The initial stage of the HHL algorithm is quantum phase estimation (QPE), utilized for eigenvalue estimation. QPE itself consists of three
subroutines: applying Hadamard gates on N-qubit clock register, applying controlled unitary transformations on the encoded b state vector achieved by the Hamiltonian
simulation and applying the inverse Quantum Fourier transform on the $N$-qubit clock register at the end, before the second part of the algorithm -- the eigenvalue inversion rotation.
Phase estimation is one of the key steps in the HHL algorithm. Efficient phase estimation of an eigenvalue can be achieved by implementing the unitary operator $e^{iA}$, which exponentiates a matrix. The number of controlled-U operations can accomplish the goal of phase estimation, which is to estimate the phase $\phi$ of the eigenvalue $v$. The middle register, also called the clock register, will be in the zero state at the beginning and end of the algorithm, but throughout the process, it will contain the binary representations of the eigenvalues. The more qubits used here, the more digits can be employed to represent the eigenvalues, leading to greater accuracy. However, using more qubits also results in a slower algorithm execution time.

$N \times N$ matrix $A$ can be expressed in terms of its eigenvectors and eigenvalues as:
    \begin{equation}
        A=\sum_{i=0}^{N-1}\psi_i |u_i\rangle \langle u_i |
    \end{equation}

\noindent Vector $b$ can also be constructed using the eigenbasis of the $A$~matrix:
    \begin{equation}
        b=\sum_{j=0}^{N-1}b_j |u_j\rangle
    \end{equation}

\noindent The solution vector $x$ can be written as~follows:
    \begin{equation}
         |x\rangle=\sum_{i=0}^{N-1}\psi_i ^{-1}b_j |u_i\rangle
    \end{equation}
\newline
\newline

Matrix $A$ must be Hermitian and of dimension $N \times N$, which means it is a square matrix and self-adjoint ($A=A^*$, equal to its own conjugate transpose). Its eigenvalues are real numbers, and it is non-singular. If it is not Hermitian, it can be modified to become~one:
\begin{center}
\begin{equation}
\begin{bmatrix}
    0 & A\\
    A^* & 0
\end{bmatrix}
\end{equation}
\end{center}
Matrix $A$ must also be $s$-sparse (where $s$ is the number of non-zero elements per row or column of the matrix) and well-conditioned. Achieving a low condition number is highly desirable, as it leads to better accuracy in numerical computation.

In the context of the HHL algorithm, Hamiltonian simulation represents a subroutine within the quantum phase estimation. The Hamiltonian represents the energy operator and Hamiltonian simulation seeks algorithms capable of efficient quantum
state time evolution implementation. Quantum Fourier Transform (QFT) represents an essential part of quantum phase estimation. Inverse QFT is needed to readout the eigenvalue. It consists of a sequence of Hadamard gates and controlled unitary rotation gates applied to N-qubits. In the quantum domain, QFT stands for basis change -- from computational to Fourier.

After the measurement, the ancilla qubit has to be in the state $\ket{1}$ for the algorithm to produce the correct solution. If the state $\ket{0}$ is output after the measurement, the solution is discarded, and the process must be repeated, which makes the algorithm probabilistic in terms of obtaining the correct solution.
An example of HHL circuit for a $2 \times 2$ matrix $A$ and a two-dimensional vector $b$ is presented in Fig.~\ref{fig:hhl}.

\subsection{Software Supply Chain Attacks}

The first attack, IIA, is based on attackers ability to modify the initial state of the victim's qubits. This can be achieved in different ways. Existing crosstalk attacks~\cite{harper2024crosstalk, ash2020analysis}, used in multi-tenant setting, have been used to manipulate victim's qubits by an attacker who executes operations on physically adjacent qubits. Another means of getting the qubits into incorrect initial state is to manipulate the user's circuit. In classical computing, well-known software supply-chain attacks~\cite{ladisa2023sok} have been studied and demonstrated, where attackers are able to modify code packages, or otherwise manipulate the supply chain of the software so that user's code is modified unbeknown to them.

\subsection{Higher Energy States}
\label{subsec:higher_energy}

The second attack, HEA, is based on higher energy states. In practical quantum computing, qubits are generally manipulated between low-energy states such as $\ket{0}$ and $\ket{1}$. The ground state, the lowest energy state of a quantum system, is commonly represented by $\ket{0}$ for a qubit. The $\ket{1}$ is the elevated energy state in the typical two-level quantum mechanical system. 
Excited states, which have higher energy than the ground state, also exist. For example, excited states with even higher energy levels such as $\ket{2}$, $\ket{3}$, and so on, can~exist. Most gate-level quantum computers only require $\ket{0}$ and $\ket{1}$, and the higher energy states are not used -- although they can still be generated.

The higher energy states are not typically used directly in quantum computations, but there is nothing that prevents users with pulse-level access to quantum computers, which is common in machines such as IBM's quantum computers, from generating such states. It is thus crucial for understanding the security of quantum systems to think about these higher energy states.

A higher energy state attack could involve an attacker manipulating a qubit to force it from its predetermined state into a higher energy state above $\ket{1}$. For superconducting quantum systems, we assume that the user has pulse-level control over qubits, which is reasonable because IBM Quantum~\cite{ibmquantum} already provides such tools such as Qiskit Pulse~\cite{qiskit2024}. With malicious experiments to obtain information from the specific qubit, one can build a custom pulse to excite a qubit to higher energy state. Previous researches have shown that higher energy state such as $\ket{2}$ is  harmful to the fidelity of quantum circuit outputs~\cite{xu2023securing}. First, the higher energy state can disturb measurement and state discriminator, which leads to wrong measure output `1' for all higher energy states. Second, the frequency of the higher energy state is different from normal states $\ket{0}$ and $\ket{1}$. This property will disable all predefined and well-calibrated gates for $\ket{0}$ and $\ket{1}$ because superconducting systems employ microwave pulses, which match the corresponding frequency of qubits, to manipulate their qubits. Another property previously discovered is that the higher energy states cannot be properly reset without using specially designed CSR gate~\cite{xu2023securing}. This means if the adversary excites qubits to higher energy states, the following user who is allocated to the same set of qubits may encounter wrong initialization issue. These can, as a consequence, prevent the quantum system from producing the correct output and become a serious security issue for superconducting quantum~systems in the cloud.

\subsection{IIA and HEA Attacks in NISQ Quantum Computer Setting}

This work evaluates the IIA and HEA attacks on Noisy Intermediate-Scale Quantum (NISQ) computers. NISQ quantum computers lack error correction and error mitigation, making them susceptible to various types of noise. This noise can affect the fidelity of the HHL algorithm, as well as the attacks themselves. The experiments in this paper were performed on NISQ quantum computers in the presence of noise, demonstrating that the attacks are effective. A detailed study of the noise's impact on IIA and HEA, however, is left as future work.

%% file: sections/threat_model.tex
\section{Threat Model}

To understand the IIA and HEA attacks, this work makes the following assumptions. We assume that the attacker does not have any special privileges and only has user-level, remote access to quantum computers. We also assume they can use tools such as Qiskit Pulse to generate custom pulses, which requires no special privileges today.

We assume that the attacker and victim share the same quantum computer, either spatially or temporally. We further assume that the attacker knows which qubits the victim is using. However, as our experimentation shows, it is not necessary to know which victim qubit is used for the ancilla, clock, or b register, since the attack on any one of them is effective.

For the IIA, we assume the attacker is able to set the victim's qubit or qubits into $\ket{1}$ state at beginning of each shot of victim's circuit. For the HEA, we assume the attacker is able to set the victim's qubit or qubits into $\ket{2}$ or other higher energy states at beginning of each shot of victim's~circuit. Existing work has already considered how to set qubit states into improper states. For example, it has shown that higher energy states are not properly reset by {\tt reset} gates~\cite{tan2023extendingdefendingattacksreset}. This work does not explore how to launch such attacks, but instead focuses on their impact on the HHL~algorithm.

%% file: sections/set_up.tex
\section{Experimental Setup}

In this section, we present the experimental setup. The experiments were performed both in simulation and on real superconducting quantum computers that were accessed remotely through the IBM Quantum service. 

\subsection{Quantum Simulators Used}

\begin{figure}[t]
    \centering
    \includegraphics[width=0.45\textwidth]{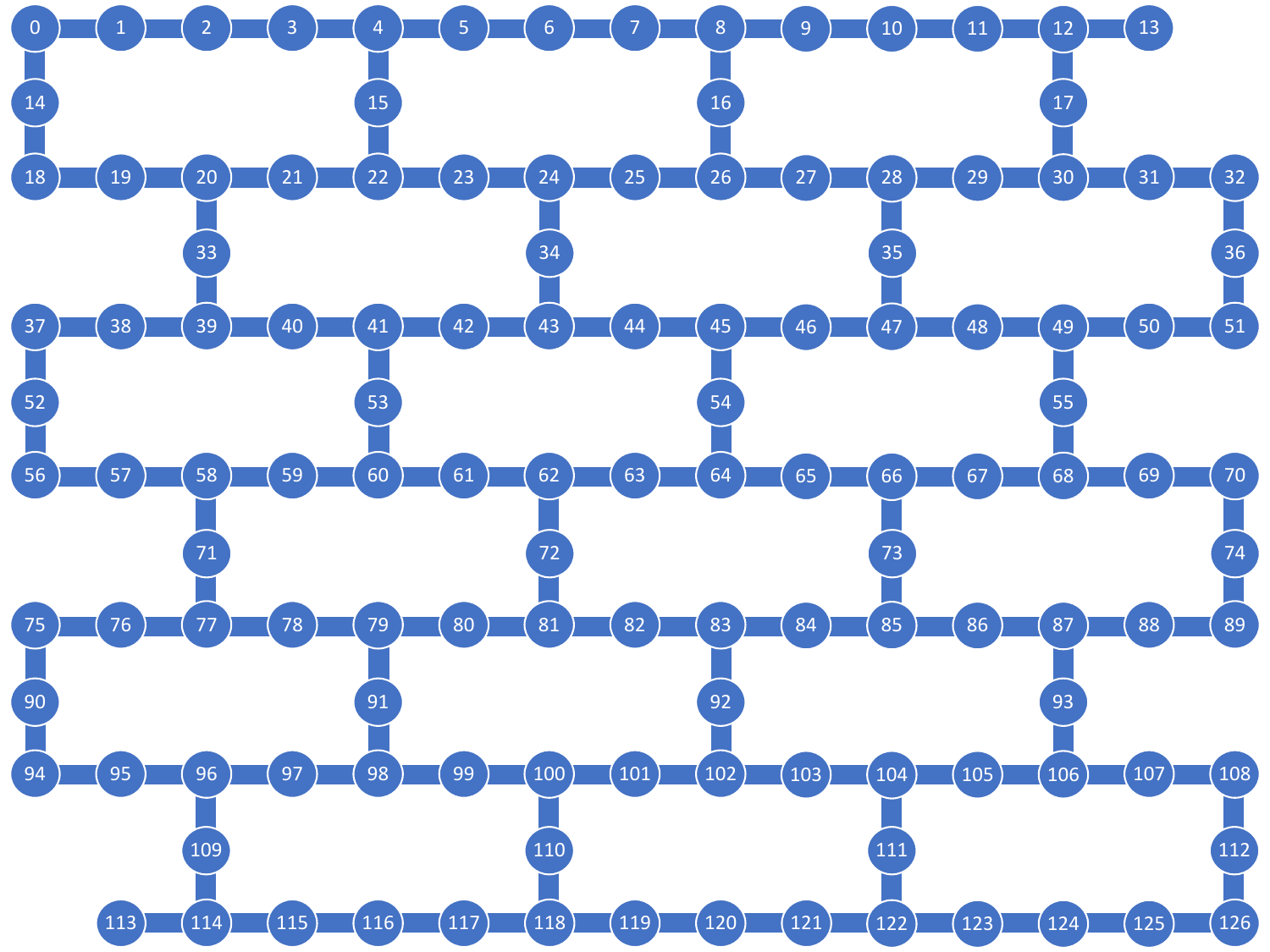}
    \caption{Topology of a IBM Quantum QPU, the Eagle r3 processor, with 127 qubits. Circles represent qubits, thick lines represent fixed couplings between the qubits.}
    \label{fig:ibm_topology}
\end{figure}

\begin{figure*}[t]
\center{\includegraphics[width=\textwidth]{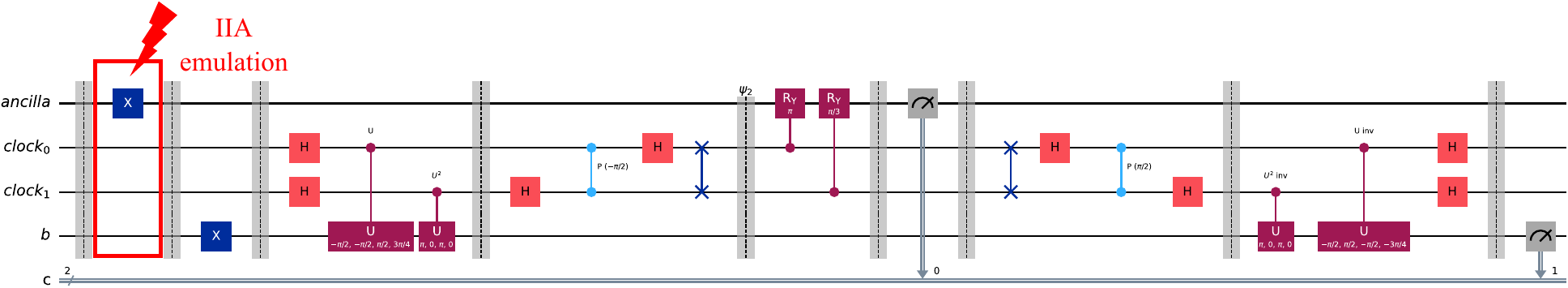}}
\caption{Emulation of Improper Initialization Attack (IIA) on HHL ancilla qubit. To emulate the attack, an additional {\tt X} gate is inserted at the beginning of the circuit in order to set the ancilla qubit to $\ket{1}$ state before the circuit executes.}
\label{fig:circuit_improper_initialization_attack}
\end{figure*}

\begin{figure*}[t]
\center{\includegraphics[width=\textwidth]{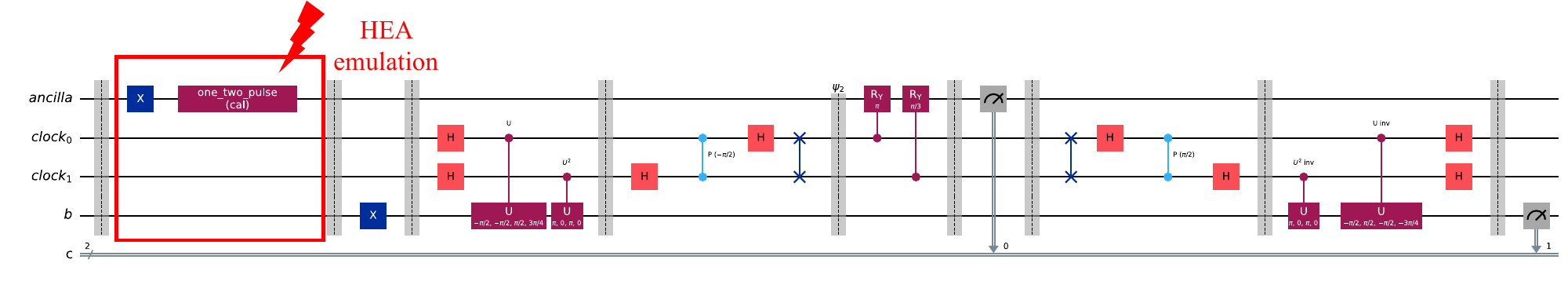}}
\caption{Emulation of Higher Energy Attack (HEA) on HHL ancilla qubit. To emulate the attack, an additional {\tt X} gate is inserted at the beginning of the circuit in order to set the ancilla qubit to $\ket{1}$ followed by a custom pulse used to excite the qubits from $\ket{1}$ to $\ket{2}$ state, i.e. the higher energy state, before the circuit executes.}
\label{fig:circuit_higher_energy_attack}
\end{figure*}

Two types of quantum simulators were used for the experiments: BasicSimulator and AerSimulator, both from IBM. BasicSimulator is part of Qiskit and is one of the ``providers''. A provider is credited with supplying external services (such as objects) to Qiskit. It can be accessed using the following class: {\tt qiskit.providers.basic\_ provider.BasicSimulator} in Qiskit. 
On the other hand, AerSimulator is part of Qiskit Aer and can be accessed using the following class: {\tt qiskit.providers.aer.backends.aerbackend.
AerBackend}.

The two simulatrs where used as-is without any modifications. In all cases, the number of shots for a quantum circuit is set to $1000$ to obtain a reasonable number of outputs for computing the output probabilities, while limiting the duration of the experiment. Since the simulators are gate-level simulators that support only $\ket{0}$ and $\ket{1}$ states, only the attacks based on IIA were evaluated on them. For HEA, real quantum hardware was used.

\subsection{Quantum Hardware Used}

In addition to the simulation, we used real quantum hardware, namely the $IBM\_brisbane$ machine, which has $127$ qubits arranged in a heavy-hexagonal layout. The physical qubits used in the experiments are qubits 0, 1, 2, and 3, or 14, since the tested HHL algorithm requires 4 qubits and the secure HHL needs 5 qubits. Larger HHL algorithms were not tested due to very noisy outputs on the real hardware we have access to.
All experiments are set to repeat for $10,000$ shots, and the optimization level was set to $0$. Each type of experiment is repeated $3$ times due to limited time available on the real hardware. The final data is the average of the corresponding 3 experiments The topology of the $IBM\_brisbane$ machine used in testing is shown in Fig.~\ref{fig:ibm_topology}.

\subsection{Generating Higher Energy States}

As mentioned in~\ref{subsec:higher_energy}, IBM Quantum provides pulse-level control over superconducting qubits through Qiskit Pulse~\cite{qiskit2024}. Although Qiskit Pulse allows users to define their custom pulses, they still need to specify the parameters for these pulses. We, and any attacker, can perform frequency sweep and Rabi experiments to obtain the parameters needed to excite a qubit into a higher energy state. These parameters are specific to each physical qubit, and the frequency sweep and Rabi experiments need to be performed on each qubit that is to be excited into higher energy states.

It should be noted that if the transpiler optimizes an input circuit and changes the layout during execution, the higher energy state may be incorrectly injected. The reason is that logical qubits are initially assigned to certain physical qubits. However, during execution, swap gates may be added by the transpiler to move the logical qubits to different physical qubits. Swap gates are ineffective when higher energy states are present, so while the logical qubits will be moved, the higher energy states will remain on the physical qubits to which they were initially assigned. As a result, we need to carefully arrange the initial layout between logical and physical qubits for each experiment to fit $IBM\_brisbane$'s topology and check the transpiled circuit to ensure we place the attack on the target qubit correctly. This is also why we use two sets of initial layouts: $[0,1,2,3]$ and $[0,1,2,14]$ for different experiments.

%% file: sections/attack.tex
\section{Circuits Used for Testing Attacks}

This work considers the two attacks: IIA and HEA. The circuits used to evaluate these attacks are shown in Fig.~\ref{fig:circuit_improper_initialization_attack} and Fig.~\ref{fig:circuit_higher_energy_attack}, respectively. The examples in the figures demonstrate the attacks on the ancilla qubits, but the same approach is taken for the attack on all other~qubits. 

For the HHL part of the testing circuits, we use the following matrix $A$:

\begin{center}
\begin{equation}
\begin{bmatrix}
    3/4 & 1/4\\
    1/4 & 3/4
\end{bmatrix}
\end{equation}
\end{center}

\noindent and vector $b$:

\begin{center}
\begin{equation}
\begin{bmatrix}
    0\\
    1
\end{bmatrix}
\end{equation}
\end{center}

\noindent The result of the HHL circuit is interpreted from the output probabilities when the b register is measured. 
The way a solution encoded in a quantum state can be compared to a classical solution vector for a particular system of linear equations is through the ratio of the squares of the magnitudes of its components.
In our case, the correct output should have a ratio of 1:9, meaning that among all the measurements when the ancilla qubit is $1$, the number of measurements of the 
b register that yield $0$ versus the number that yield $1$ should be in the ratio 1:9.

For testing the IIA attack, an additional {\tt X} gate is inserted at the beginning of the circuit in order to set the target qubit to $\ket{1}$ state before the circuit executes. This is to emulate the attack only. For testing the HEA attack, an additional {\tt X} gate is inserted at the beginning of the circuit in order to set the target qubit to $\ket{1}$ followed by a custom pulse used to excite the qubits from $\ket{1}$ to $\ket{2}$ state, i.e. the higher energy state, before the circuit executes. This is also to emulate the attack only. In practice, attackers could use different schemes to achieve the IIA and HEA.

In the experiments, we have limited the testing to HHL circuits with 4 qubits (ancilla, b, and two clock qubits). Larger HHL circuits do not work correctly on the current hardware available to us due to noise in the quantum computers; thus, testing of larger HHL circuits is left as future work.

%% file: sections/defense.tex
\section{Defenses}

Before presenting evaluation of the attacks, we discuss the ideas and design of the defenses. Subsequently, both the attacks and the defenses are evaluated together.

\subsection{Defense Idea}
\label{sec_defense_idea}

The goal of our work is to detect when an attack has occurred and allow the user to determine from the output of the quantum circuit if there is an attack (and results should be ignored) or if there was no attack.
The approach to detecting the presence of attacks involves incorporating additional measurements for the different qubits used by the HHL algorithm: ancilla, clock, and
b qubits. For each qubit we need to detect if it may have been attacked through IIA or HEA.

To detect IIA, the intuition is that we need to confirm if a qubit was initialized into $\ket{0}$ state (no attack) or $\ket{1}$ (attack). This can be done, for example, by directly measuring the qubit at beginning of the execution and checking it. An interesting property of the clock qubits used in HHL, is that HHL performs uncomputation the clock qubits and the final state of the qubits is the same as the initial state -- in this case the measurement and checking of the clock qubits can be delayed to the end of the circuit, which is HHL specific optimization.

To differentiate between IIA and HEA, we leverage the property of higher-energy states: the predefined quantum gates are not effective in presence of higher energy states. Both improper initialization and higher-energy states results in the qubits being set into initial states that would be both measured as '1'. However, improper initialization sets the qubit(s) into $\ket{1}$ state that is modified when different quantum gates are applied, while for higher-energy states, the qubit(s) are set into $\ket{2}$ or higher states, that are not readily affected by the quantum gates -- no matter what gates are applied to the qubits, the measurement of the qubit(s) will always be '1' (until the qubits begin to decay). When possible, the defense aims to differentiate if the quantum states are or are not affected by the HHL algorithms gates, thus pointing to the type of attack.

\subsection{Defenses Circuit}

\begin{figure*}[t]
\center{\includegraphics[width=\textwidth]{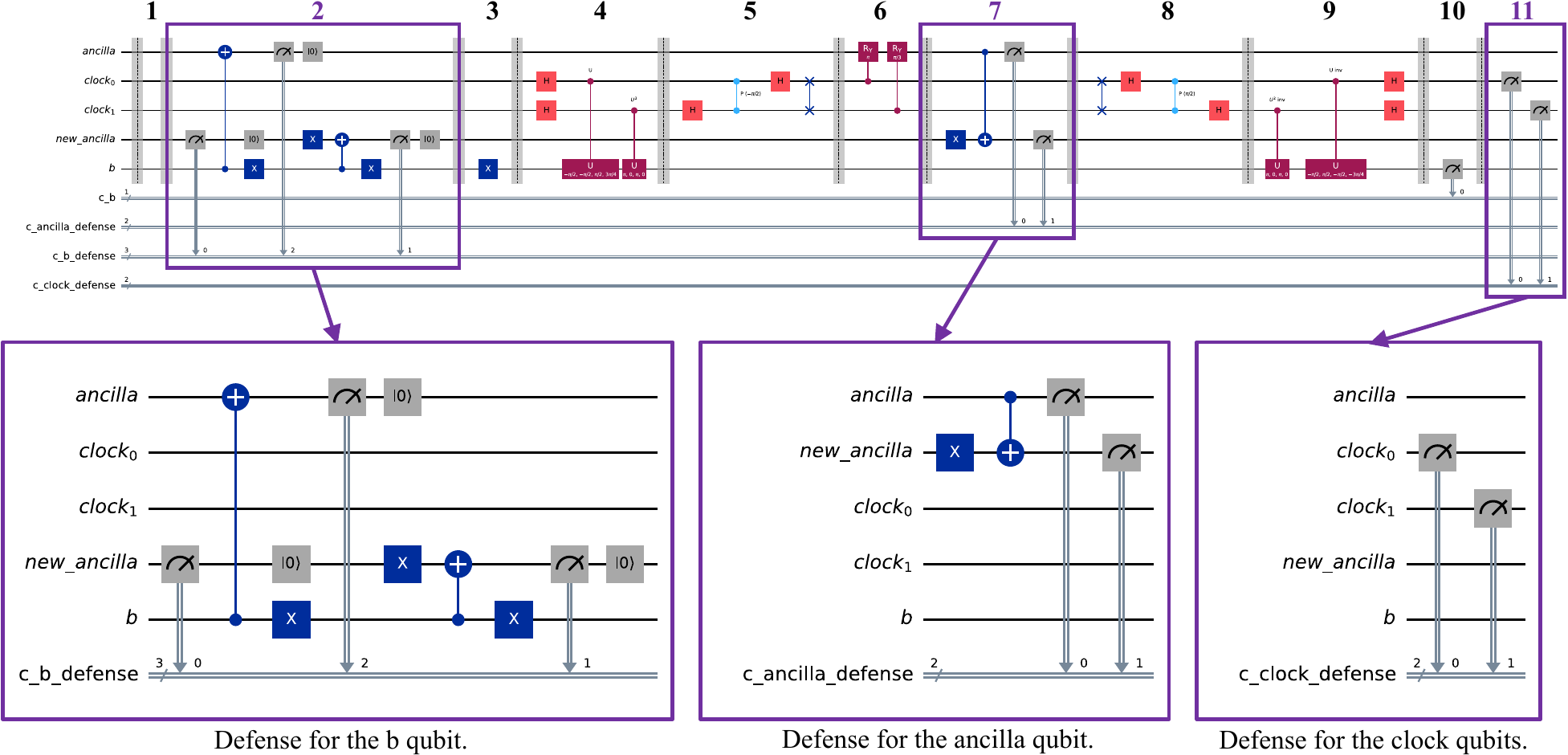}}
\caption{HHL circuit including our defenses. Part 2 serves as a defense for the b qubit,
Part 7 serves as a defense for the ancilla qubit, while
Part 11 serves as a defense for the clock qubits.}
\label{fig_defense_circuit}
\end{figure*}

We implement our ideas as a modified HHL circuit, with the defense strategies incorporated against both IIA and HEA on the HHL algorithm using only one extra new ancilla qubit. This approach not only protects all qubits in the HHL algorithm but also enables the detection of the attack type. The defenses are integrated into the HHL algorithm's circuit as shown in Fig.~\ref{fig_defense_circuit}. THis is an example where HHL is configured for a 2x2 matrix $A$ and a two-dimensional vector $b$. This defense can be easily scaled to larger HHL circuits in future work.

In Fig.~\ref{fig_defense_circuit}, we clearly divide the circuit into 11 parts based on their roles. Part 1 is where a potential attack may occur on the physical qubits of the superconducting quantum computer. Part 2 is a specially inserted defense for the b qubit, which will be explained later. Part 3 (state preparation), Part 4 (QPE), Part 5 (inverse QFT), and Part 6 (controlled rotation) are all parts of the original HHL algorithm, and are unmodified. We adjust Part 7, which previously only involved the measurement of the ancilla qubit, to serve as the defense for the ancilla qubit. The following, Part 8 (QFT), Part 9 (inverse QPE), and Part 10 (b measurement), are also directly taken from the remaining HHL algorithm. The additional Part 11 is the defense for clock qubits.

The different measurements performed during circuit execution in Parts 2, 7, and 11 are stored in classical registers. The output of these registers can be used by the user to determine if any attack occurred. As detailed later, when the measurement output from the {\tt b\_defense} is $00$, the measurement from the {\tt ancilla\_defense} is $10$, the measurement from the {\tt clock\_defense} is $00$, and the measurement of b qubit is $1$, we know that the HHL algorithm is safe and convergent. If in the same case the b qubit is $0$ then the HHL has not yet converged, but there was no attack. In All other cases there has been an attack.

\begin{table*}[t]
    \centering
    \caption{\small Determining if an attack occured on the ancilla and new ancilla qubits. The {\tt c\_ancilla\_defense} output is measured in Part 7 in Fig.~\ref{fig_defense_circuit}.}
    \small
    \begin{tabular}{c|c}
    \hline
    \textbf{Attack Type} & \textbf{Expected} {\tt c\_ancilla\_defense} \textbf{Output} \\ \hline
    \textbf{No attack, HHL converges} & {\tt 10}\\ \hline
    \textbf{No attack, HHL continues to update} & {\tt 01}\\ \hline
    \textbf{HEA on ancilla} & {\tt 11} \\ \hline
    \textbf{HEA on new ancilla} & {\tt 01} or {\tt 11} \\ \hline
    \textbf{HEA on both ancilla and new ancilla} & {\tt 11} \\ \hline
    \end{tabular}
    \label{tab:defense_ancilla}
\end{table*}

\begin{table*}[t]
    \centering
    \caption{\small Determining if an attack occured on the b qubit. The {\tt c\_b\_defense} output is measured in Part 2 in Fig.~\ref{fig_defense_circuit}.}
    \small
    \begin{tabular}{c|c}
    \hline
    \textbf{Attack Type} & \textbf{Expected} {\tt c\_b\_defense} \textbf{Output} \\ \hline
    \textbf{No attack} & {\tt 000} \\ \hline
    \textbf{HEA on b} & {\tt 010} \\ \hline
    \textbf{IIA on b} & {\tt 011} \\ \hline
    \textbf{HEA on ancilla} & {\tt 001} \\ \hline
    \textbf{IIA on ancilla} & {\tt 001} \\ \hline   
    \textbf{HEA on new ancilla} & {\tt 110} \\ \hline
    \textbf{IIA on new ancilla} & {\tt 100} \\ \hline
    \textbf{HEA on ancilla and b} & {\tt 011} \\ \hline
    \textbf{IIA on ancilla and b} & {\tt 010} \\ \hline
    \textbf{HEA on ancilla and new ancilla} & {\tt 111} \\ \hline
    \textbf{IIA on ancilla and new ancilla} & {\tt 101} \\ \hline
    \textbf{HEA on new ancilla and b} & {\tt 110} \\ \hline
    \textbf{IIA on new ancilla and b} & {\tt 111} \\ \hline
    \textbf{HEA on ancilla, new ancilla and b} & {\tt 111} \\ \hline
    \textbf{IIA on ancilla, new ancilla and b} & {\tt 110} \\ \hline
    \end{tabular}
    \label{tab:defense_b}
\end{table*}

\begin{table*}[t]
    \centering
    \caption{\small Determining if an attack occured on the clock qubits. The {\tt c\_clock\_defense} output is measured in Part 11.}
    \small
    \begin{tabular}{c|c}
    \hline
    \textbf{Attack Type} & \textbf{Expected} {\tt c\_clock\_defense} \textbf{Output} \\ \hline
    \textbf{No attack} & {\tt 00} \\ \hline
    \textbf{HEA on clock0} & {\tt 10} \\ \hline
    \textbf{IIA on clock0} & {\tt 10} \\ \hline
    \textbf{HEA on clock1} & {\tt 01} \\ \hline
    \textbf{IIA on clock1} & {\tt 01} \\ \hline
    \textbf{HEA on clock0 and clock1} & {\tt 11} \\ \hline
    \textbf{IIA on clock0 and clock1} & {\tt 11} \\ \hline
    \end{tabular}
    \label{tab:defense_clock}
\end{table*}

\subsection{Details of Defense for the ancilla Qubit}
\label{subsec:ancilla_defense}

We explain the ancilla defense first because this protects both the ancilla qubit and the new ancilla qubit, which are used in the defense for the b qubit. It is important to ensure that the user has protected the ancilla qubit, even though this defense is positioned in the middle of the circuit. For the defense of the ancilla from attack, a new ancilla qubit was added to the original HHL circuit as shown bottom-middle of Fig.~\ref{fig_defense_circuit}. Additionally, an {\tt X} gate was applied to the new ancilla qubit, and a {\tt CNOT} gate was added with the ancilla qubit as the control qubit and the new ancilla qubit as the target qubit. Finally, two measurement gates were added: one for the ancilla qubit and one for the new ancilla qubit. The measurement outputs are stored in the classical register {\tt c\_ancilla\_defense}. The bit on the left of {\tt c\_ancilla\_defense} is the measurement of the ancilla qubit while that on the right is for the new ancilla qubit.
Normally, these qubits will be measured as $10$ when HHL converges or $01$ when further updates are required, while under attack they will be measured as $11$ because HEA results in a '1' readout and disables the {\tt CNOT} gate.
Consequently, the error introduced by IIA or HEA on the ancilla qubit can result in unexpected outputs from {\tt c\_ancilla\_defense}.
Table~\ref{tab:defense_ancilla} shows how the output of {\tt c\_ancilla\_defense} can be used to determine if there was an attack, and what type of attack. While output $01$ seems ambiguous, by using {\tt c\_b\_defense} discussed later we can fully determine if $01$ means no attack or HEA on new ancilla.

\subsection{Details of Defense for the b Qubit}

Bottom-left of Fig.~\ref{fig_defense_circuit} shows details of the defense for the b qubit using two ancilla qubits.
We apply {\tt X} gates to the b qubit twice. Initially, the qubit should be in $\ket{0}$ state. If the b qubit is not under attack, two {\tt X} gates (which are analogous to NOT gates in classical computers) will cancel each other out and the state of the b qubit will be $\ket{0}$ and can be used by the remainder of the algorithm as normal. To detect a possible attack, the b qubit is entangled with the ancilla and new\_ancilla qubits.

This efficient defense can distinguish the state of the b qubit simply by applying {\tt CNOT}, {\tt X}, and {\tt Reset} gates, along with measurements on two ancilla qubits stored in {\tt c\_b\_defense}, thereby avoiding direct measurement of b itself. The leftmost bit is the first measurement of the new ancilla qubit. The output of the second new ancilla measurement is stored in the second bit while that on the right is for the ancilla qubit. The {\tt Reset} gates in this defense can also protect ancilla and new ancilla qubits from IIA, which is why we do not mention IIA in~\ref{tab:defense_ancilla}.

Table~\ref{tab:defense_b} outlines how {\tt c\_b\_defense} can be used to determine if there was an attack.

Some attack types may seem ambiguous due to identical expected {\tt c\_b\_defense} output. However, by combining both {\tt c\_b\_defense} and {\tt c\_ancilla\_defense}, we can distinguish these attack types to a certain degree. For example, HEA on ancilla, IIA on ancilla both yield the same expected {\tt c\_b\_defense} output $10$. However, when {\tt c\_ancilla\_defense} = $11$, it indicates that an HEA has occurred on the ancilla qubit. Regardless, only one output of {\tt c\_b\_defense}, i.e. $000$, indicates no attack.

\subsection{Details of Defense for the Clock Qubits}

As shown in bottom-right of Fig.~\ref{fig_defense_circuit}, the defense for the clock qubits consists of adding two measurement gates at the end of the circuit, one for each clock qubit. The clock measurements are stored in {\tt c\_clock\_defense}. The bit on the left of {\tt c\_clock\_defense} means the output of clock0, and the one on the right is the output of clock1. 
The uncomputation part of HHL algorithm, as shown in part 8 and 9 in Fig.~\ref{fig_defense_circuit}, will recover clock qubits back to their initial states. As a result, if the circuit is not under attack, the output on both clock qubits should be equal to the initial state $\ket{0}$, which is measured as '0'.
All other measurement results of the clock qubits (e.g., some amplitude ratio between $\ket{0}$ and $\ket{1}$, or state $\ket{1}$ as the output) indicate that these qubits are under attack. For IIA, the qubits are initially attacked and set to $\ket{1}$ and so the final measurement will be '1' since the uncomputation returns the qubits to their initial states. For HEA, the any gates in the circuit are not effective on the higher energy states, which are measured as '1', and the final measurement will be '1'. For both IIA and HEA, the final measurement is '1' instead of '0' for the attacked~qubit.

Table~\ref{tab:defense_clock} summarizes how output of the {\tt c\_clock\_defense} can be used in determining if an attack occurred on the clock qubits. The {\tt c\_clock\_defense} output is measured in Part 11 of Fig.~\ref{fig_defense_circuit}. When there is no attack, the expected measurement of the two clock qubits, saved in the {\tt c\_clock\_defense} register is $00$. Meanwhile, in any other case, the output is other than $00$ indicating an attack has happened.

\subsection{Combined Defense and Attack Detection}

All the defenses combined in the HHL circuits are again shown in Fig.~\ref{fig_defense_circuit}. To detect if an attack has affected the circuit, the user need to read out the concatentaed value of {\tt c\_ancilla\_defense} (2 bits), {\tt c\_b\_defense} (3 bits) and {\tt c\_clock\_defense} (2 bits) registers. The results is a 7-bit value. Value of $10 000 00$ indicates no attack, HHL converges. Value of $01 000 00$ indicates no attack, HHL continues to update. All other values indicate an attack and computation results should be discarded.

%% file: sections/simulation.tex
\section{Evaluation in Simulation}
\label{sec:simulator}

\begin{figure}[t]
 \centering
 \begin{minipage}{0.48\linewidth}
      \includegraphics[width=\textwidth]{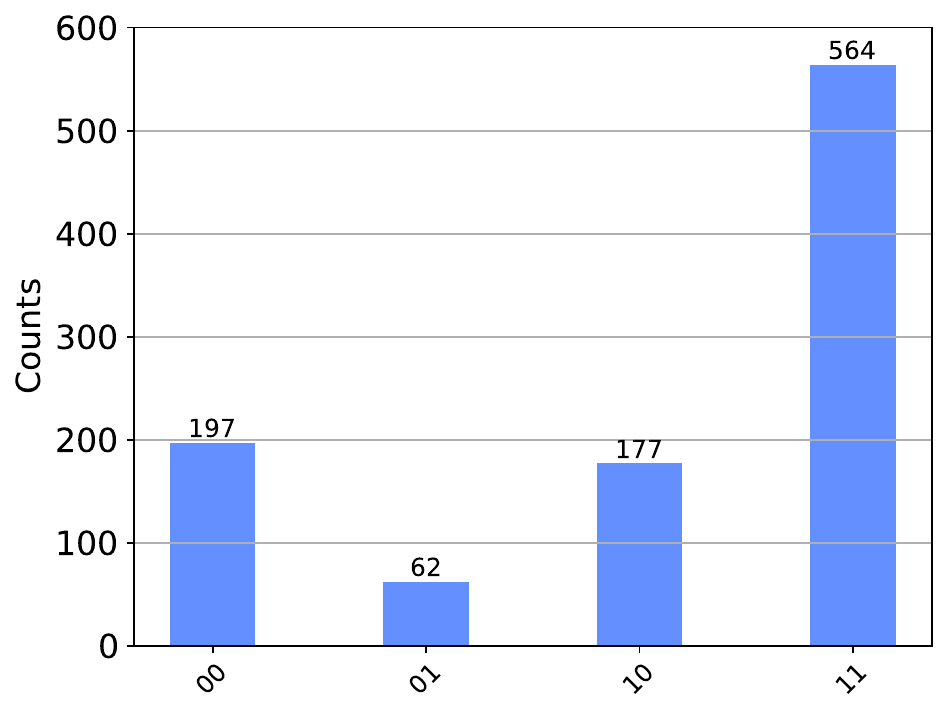}
    \label{subfig:hhl_basic}
 \end{minipage}
  \begin{minipage}{0.48\linewidth}
      \includegraphics[width=\textwidth]{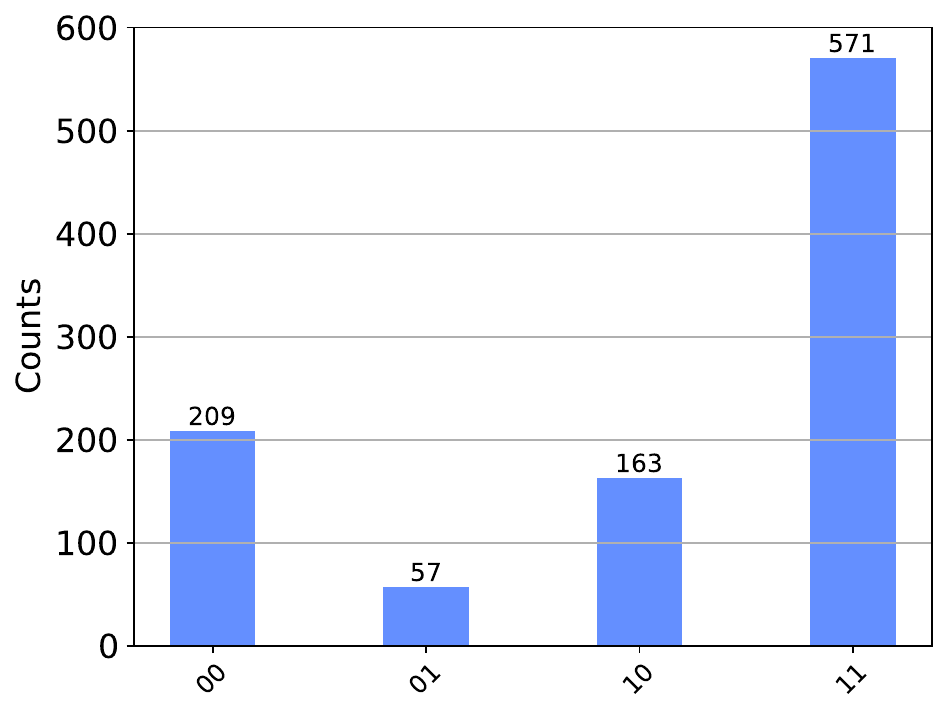}
    \label{subfig:hhl_aer}
 \end{minipage}
 \caption{Comparison between baseline results of running HHL on the BasicSimulator (left) and the AerSimulator (right).}
 \label{fig_baseline}
\end{figure}

\begin{figure}[t]
 \centering
 \begin{minipage}{0.48\linewidth}
      \includegraphics[width=\textwidth]{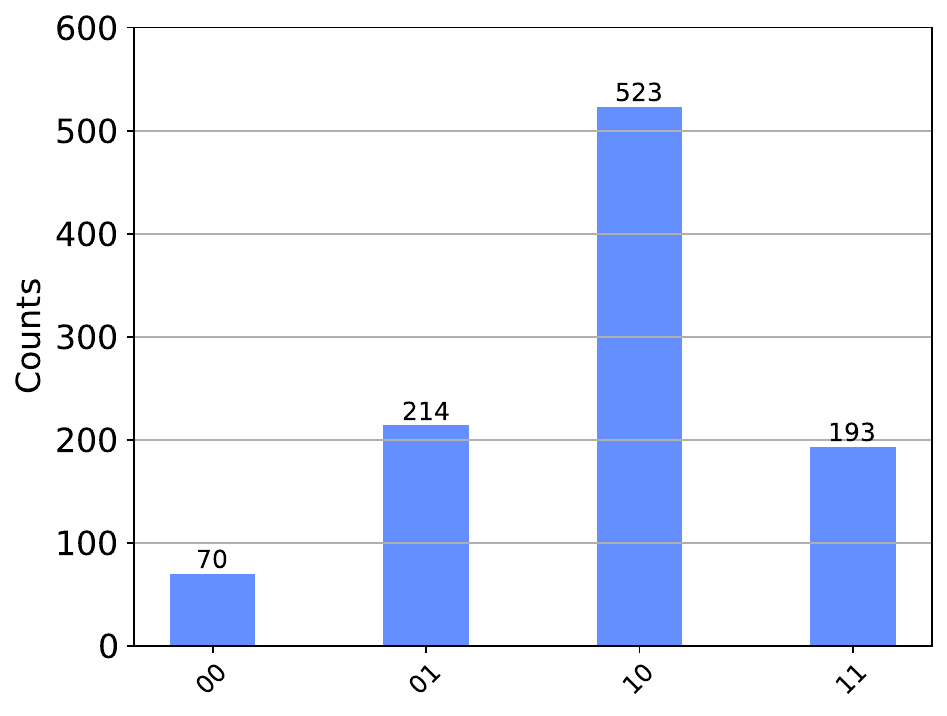}
    \label{subfig:hhl_iia_ancilla_basic}
 \end{minipage}
  \begin{minipage}{0.48\linewidth}
      \includegraphics[width=\textwidth]{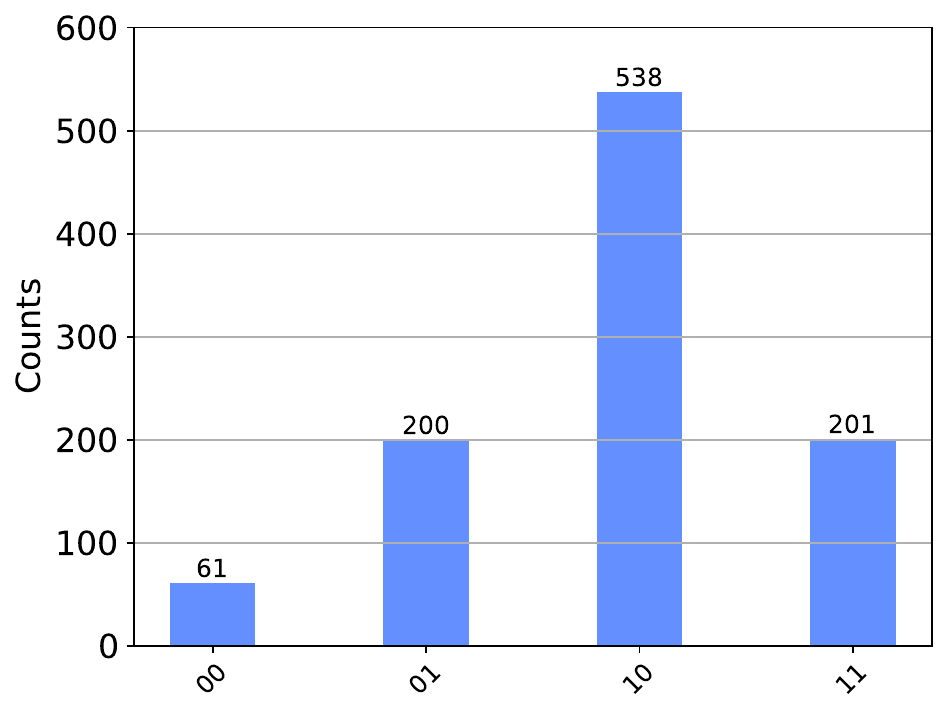}
    \label{subfig:hhl_iia_ancilla_aer}
 \end{minipage}
 \caption{Comparison of results of the IIA on the ancilla qubit tested on BasicSimulator (left) and the AerSimulator~(right).}
 \label{fig_iia}
\end{figure}

\begin{figure}[th!]
 \centering
 \begin{minipage}{0.48\linewidth}
      \includegraphics[width=\textwidth]{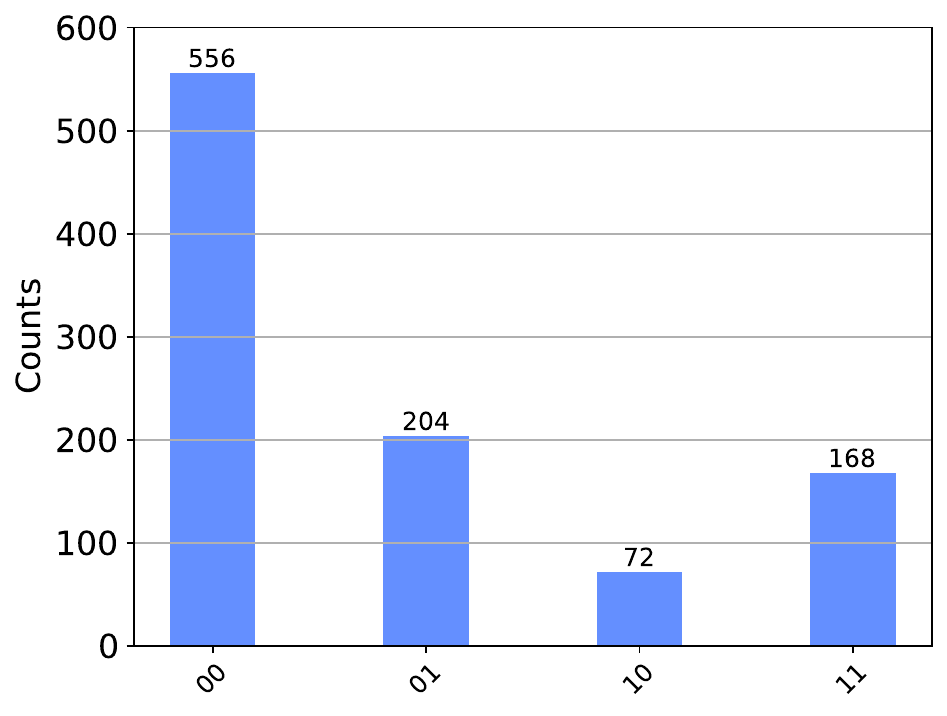}
      \subcaption{IIA on the first clock qubit.}
      \label{subfig:hhl_iia_clock0_basic}
 \end{minipage}
 \begin{minipage}{0.48\linewidth}
      \includegraphics[width=\textwidth]{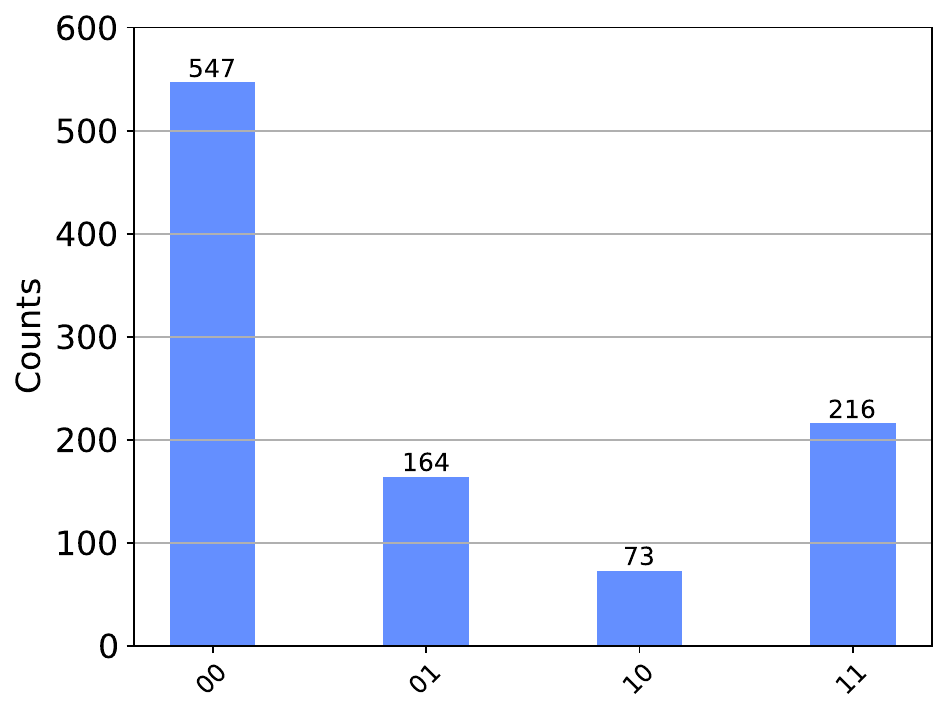}
      \label{subfig:hhl_iia_clock0_aer}
 \end{minipage}
 \begin{minipage}{0.48\linewidth}
      \includegraphics[width=\textwidth]{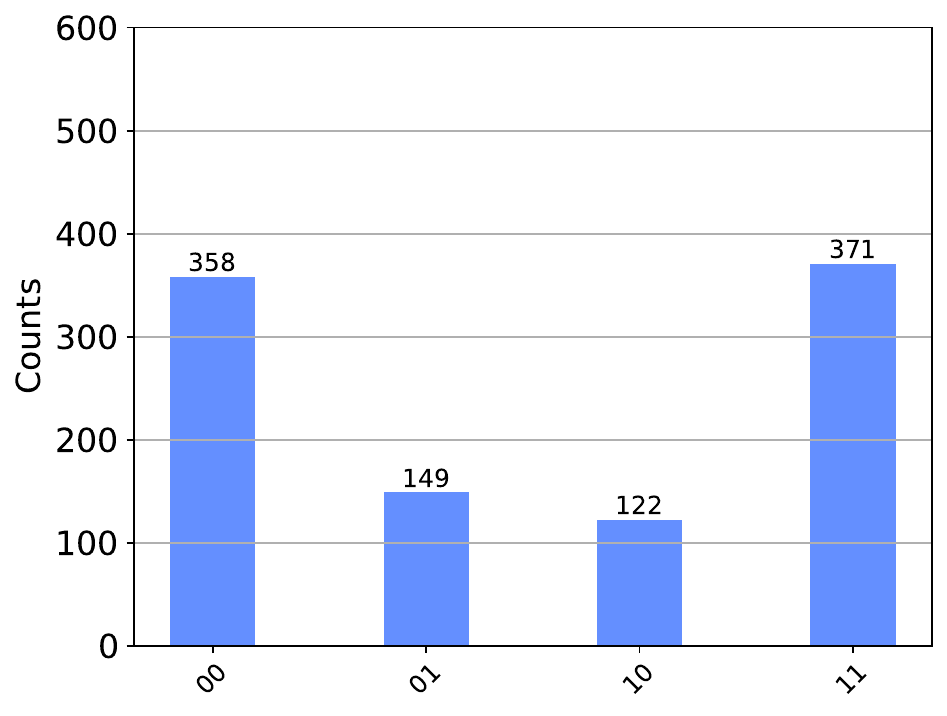}
      \subcaption{IIA on the second clock qubit.}  
      \label{subfig:hhl_iia_clock1_basic}
 \end{minipage}
 \begin{minipage}{0.48\linewidth}
      \includegraphics[width=\textwidth]{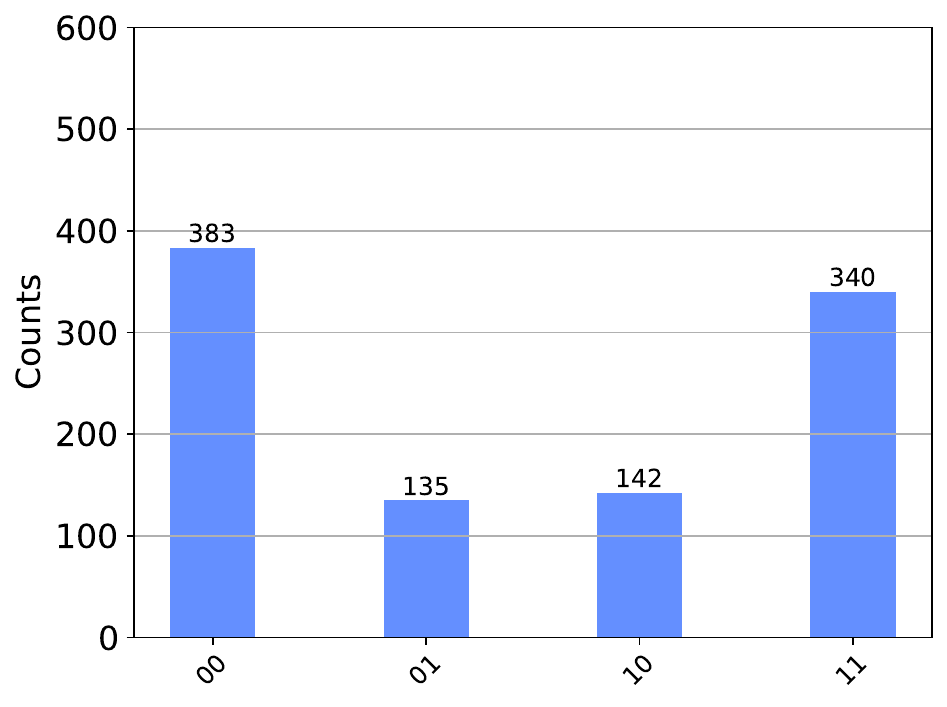}
      \label{subfig:hhl_iia_clock1_aer}
 \end{minipage}
 \begin{minipage}{0.48\linewidth}
      \includegraphics[width=\textwidth]{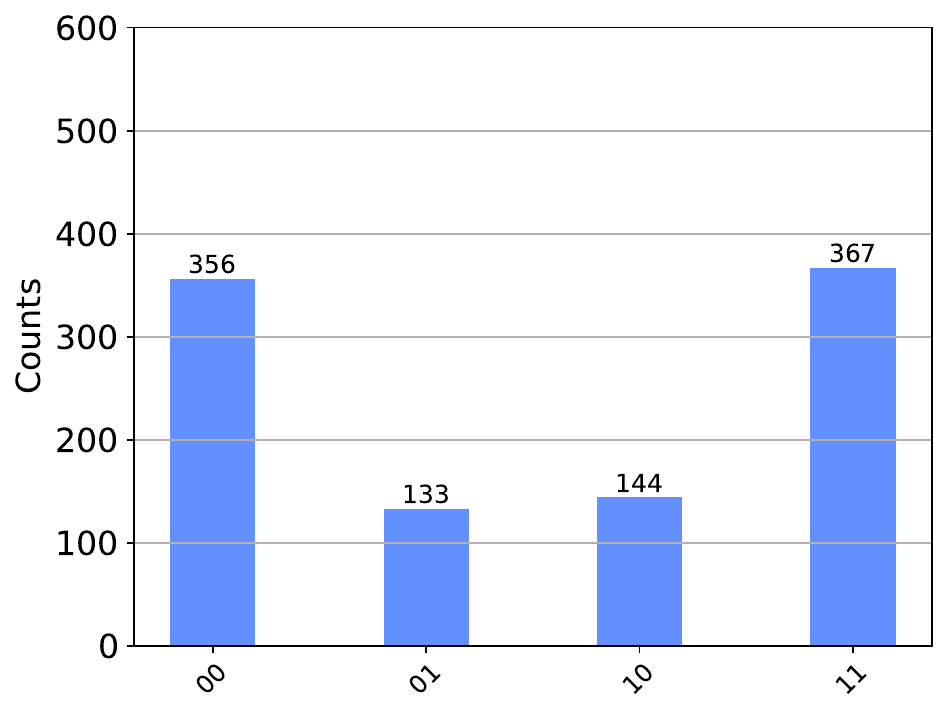}
      \subcaption{IIA on both clock qubits.}   
      \label{subfig:hhl_iia_clock01_basic}
 \end{minipage}
  \begin{minipage}{0.48\linewidth}
      \includegraphics[width=\textwidth]{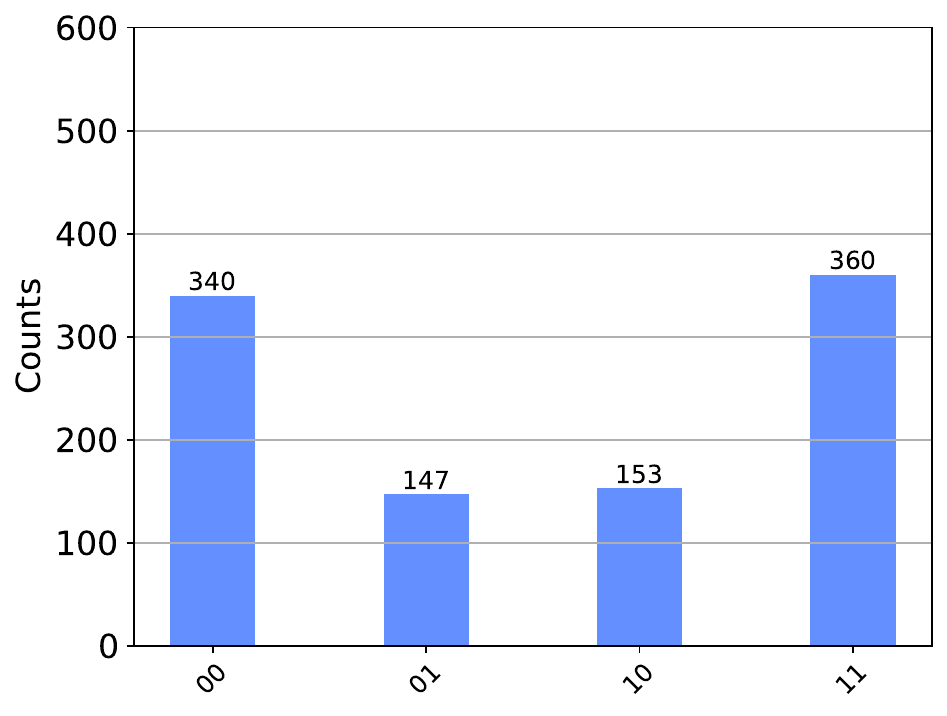}
      \label{subfig:hhl_iia_clock01_aer}
 \end{minipage}
 \caption{Comparison of results of the IIA on the two clock qubits tested on BasicSimulator (left) and the AerSimulator (right).}
 \label{fig_clocks}
\end{figure}

In this section, we present results from testing in simulation. We focus on the two simulators available in Qiskit, as discussed earlier.

\subsection{Baseline HHL Outputs}

Fig.~\ref{fig_baseline} shows the result from the BasicSimulator and the AerSimulator. The BasicSimulator yields a result of 1:9.0968, while the AerSimulator produces 1:10.0175, both of which closely approximate the classical solution of 1:9. Note that both ancilla and b qubits are measured, resulting in four possible states shown in figures: 00, 01, 10, and 11. The first bit is b and second bit is ancilla. The HHL output is only considered correct when ancilla is measured to be 1. Thus, when computing the ratios, we compute the ratio of the state 01 to state 11.

\subsection{HHL Outputs Under Attack}

In the following text we report the results of the IIA on the ancilla qubit, clock qubit, b qubit.

\subsubsection{IIA on Ancilla Qubit}

Fig.~\ref{fig_iia} shows that for the IIA on the ancilla qubit, the result from the BasicSimulator is 1:0.9019, while the result from the AerSimulator is 1:1.0050, which clearly shows that the attack worsened the result by 9 times.

\begin{figure}[t]
 \centering
 \begin{minipage}{0.48\linewidth}
      \includegraphics[width=\textwidth]{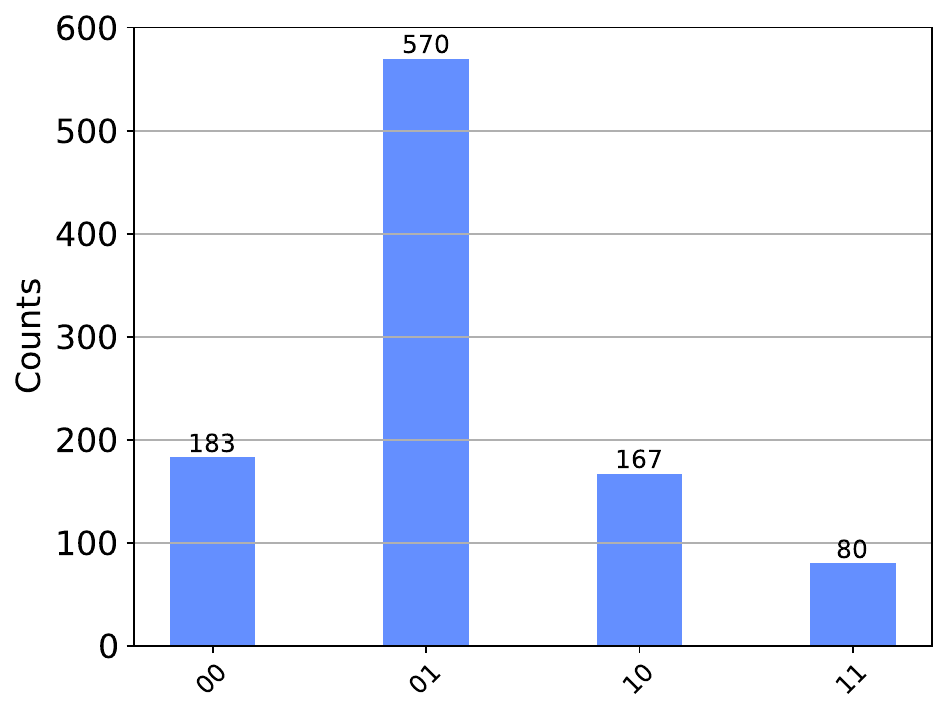}
    \label{subfig:hhl_iia_b_basic}
 \end{minipage}
  \begin{minipage}{0.48\linewidth}
      \includegraphics[width=\textwidth]{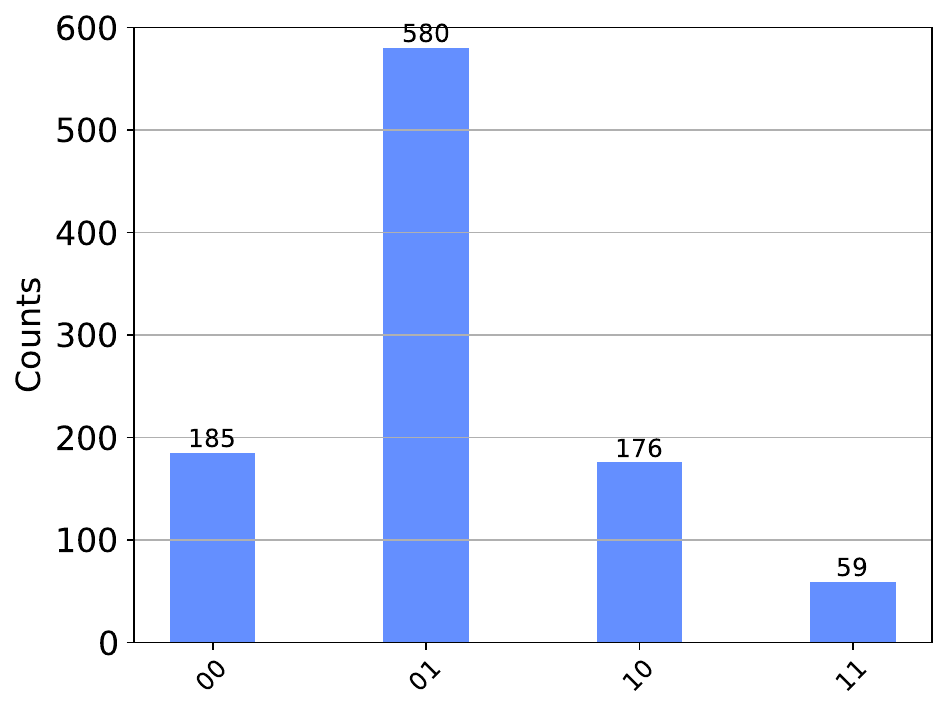}
    \label{subfig:hhl_iia_b_aer}
 \end{minipage}
 \caption{Comparison between the BasicSimulator and the AerSimulator for the IIA on the b qubit.}
 \label{fig_b}
\end{figure}

\begin{table}[t]
    \centering
    \caption{Variational distances of HHL outputs under IIA on BasisSimulator and AerSimulator.}
    \begin{tabular}{c|c|c}
    \hline
    \textbf{Victim qubit} & \textbf{BasicSimulator} & \textbf{AerSimulator} \\ \hline
    \textbf{no attack} & 0 & 0 \\ \hline
    \textbf{ancilla} & 0.4980 & 0.5180 \\ \hline
    \textbf{clock0} & 0.5010 & 0.4450 \\ \hline
    \textbf{clock1} & 0.2489 & 0.2520 \\ \hline
    \textbf{clock0 and clock1} & 0.2300 & 0.2210 \\ \hline
    \textbf{b} & 0.5080 & 0.5360 \\ \hline
    \end{tabular}
    \label{tab:improper_initialization_attack_simulator}
\end{table}

\subsubsection{IIA on Clock Qubits}

Fig.~\ref{fig_clocks} shows that for the IIA on the clock0 qubit, the result from the BasicSimulator is 1:0.8235, while the result from the AerSimulator is 1:1.3171. For the IIA on the clock1 qubit, the result from the BasicSimulator is 1:2.4899, while the result from the AerSimulator is 1:2.5185. For the IIA on both clock qubits, the result from the BasicSimulator is 1:2.7594, while the result from the AerSimulator is 1:2.4490.
The attack worsened the result by 4 times.

\subsubsection{IIA on b Qubit}

Fig.~\ref{fig_b} shows that for the IIA on the b qubit, the result from the BasicSimulator is 1:0.1403, while the result from the AerSimulator is 1:0.1017. The attack worsened the result by 65 times.

\subsection{Summary of the Effectiveness of the Attacks}

To compare the results without and with attack, we used the variational distance metric.
Table~\ref{tab:improper_initialization_attack_simulator} shows the variational distance between original HHL probability distribution and IIA probability distributions under the different attacks on ancilla, clock, and b qubits. It can be seen that attacking any of the qubits results in significant variational distance. Interestingly, attacking clock1 or both clock0 and clock1 has less impact than attacking the other~qubits. Nevertheless, we can surmise that attacking any one qubit is sufficient to generate incorrect results.

%% file: sections/hardware_attack.tex
\section{Evaluation on Quantum Hardware}

In this section we present results from testing the attacks on real hardware. We used the $IBM\_brisbane$ machine.

\begin{table}[t]
    \centering
    \caption{Ratio and variational distances of HHL outputs under HEA on $IBM\_brisbane$.}
    \begin{tabular}{c|c|c}
    \hline
    \textbf{Victim Qubit} & \textbf{Ratio} & \textbf{Variational Distance} \\ \hline
    \textbf{No attack} & 1 : 1.1759 & 0 \\ \hline
    \textbf{ancilla} & 1 : 0.8095 & 0.1863 \\ \hline
    \textbf{clock0} & 1 : 0.6849 & 0.1002 \\ \hline
    \textbf{clock1} & 1 : 1.8335 & 0.0713 \\ \hline
    \textbf{b} & 1 : 2.7800 & 0.2065 \\ \hline
    \textbf{Attack 4 qubits} & 1 : 1.2169 & 0.1191 \\ \hline
    \end{tabular}
    \label{tab:higher_energy_attack_hardware}
\end{table}

\begin{table}[t]
    \centering
    \caption{Ratio and variational distances of HHL outputs under IIA on $IBM\_brisbane$.}
    \begin{tabular}{c|c|c}
    \hline
    \textbf{Victim Qubit} & \textbf{Ratio} & \textbf{Variational Distance} \\ \hline
    \textbf{No attack} & 1 : 1.1759 & 0 \\ \hline
    \textbf{ancilla} & 1 : 0.8910 & 0.1048 \\ \hline
    \textbf{clock0} & 1 : 1.2912 & 0.1125 \\ \hline
    \textbf{clock1} & 1 : 1.1868 & 0.0923 \\ \hline
    \textbf{b} & 1 : 1.0105 & 0.0616 \\ \hline
    \textbf{Attack 4 qubits} & 1 : 1.3788 & 0.1388 \\ \hline
    \end{tabular}
    \label{tab:improper_initialization_attack_hardware}
\end{table}

\subsection{Baseline HHL Outputs}

\begin{figure}[t]
 \centering
 \includegraphics[width=0.24\textwidth]{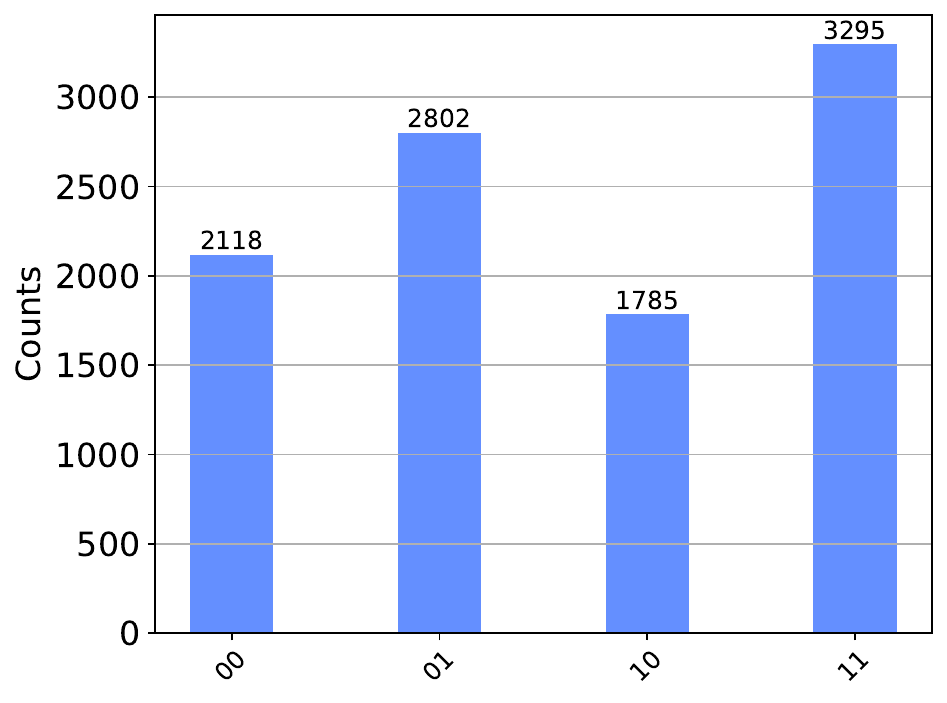}
 \caption{Baseline HHL outputs testing HHL without attacks when using qubit 0,1,2,3 on \textit{IBM\_brisbane}. The graph represents the average output of three tests.}
 \label{fig:HHL}
\end{figure}

We again use the original 2x2 HHL algorithm mentioned in Sec.~\ref{sec:simulator}. The average ratio of the three repeated HHL experiments in Fig.~\ref{fig:HHL} on $IBM\_brisbane$ machine is 1:1.759. It can be immediately seen that even without the attack, due to the noisy nature of the current hardware, the output ratio is much worst than in simulation.  

\subsection{HHL Outputs Under Attack}

In the following text we report the results of the HEA on the ancilla qubit, clock qubits, and b qubit.

\begin{figure*}
    \centering
    \begin{subfigure}[b]{0.24\linewidth}
    \centering
    \includegraphics[width=\linewidth]{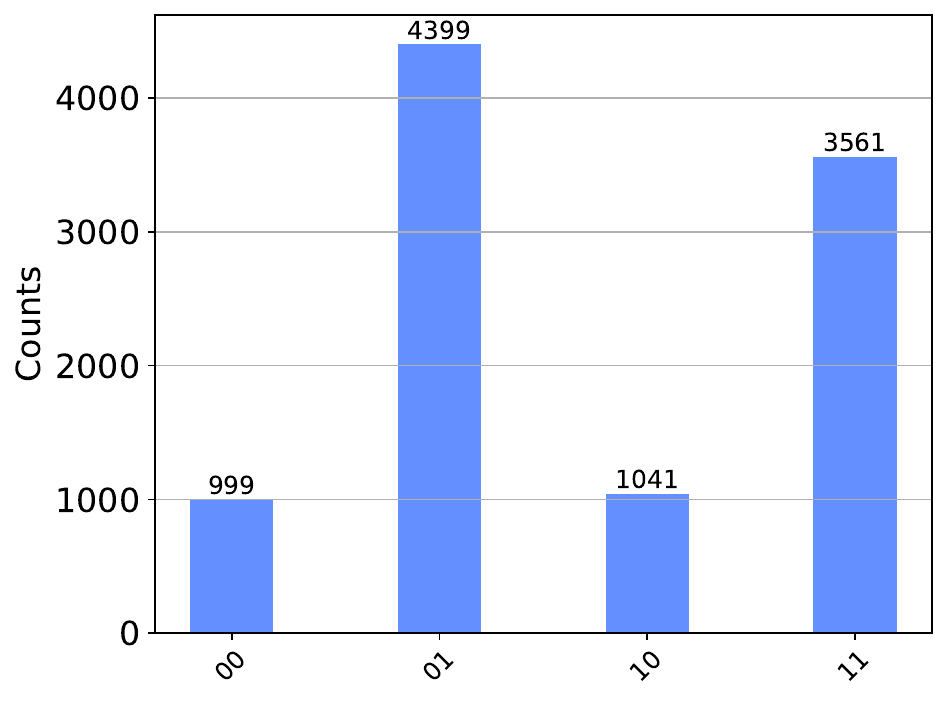}
    \caption{HEA on HHL ancilla using qubit 0,1,2,3 on \textit{IBM\_brisbane}.}
    \label{subfig:attack_a_higher_energy}
    \end{subfigure}
    \hfill
    \begin{subfigure}[b]{0.24\linewidth}
    \centering
    \includegraphics[width=\linewidth]{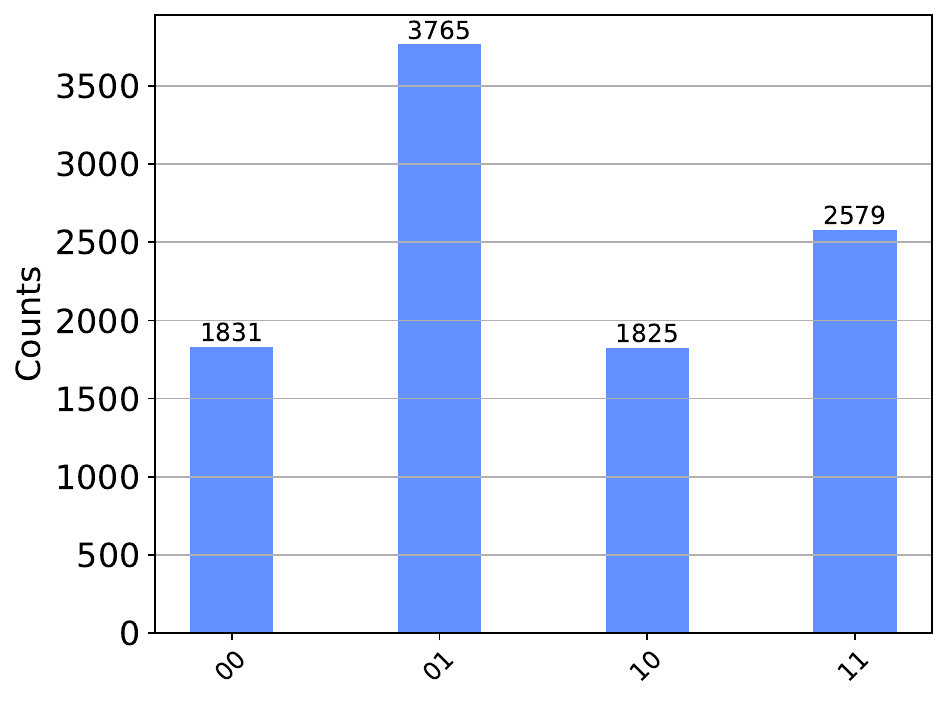}
    \caption{HEA on HHL clock0 using qubit 0,1,2,3 on \textit{IBM\_brisbane}.}
    \label{subfig:attack_clock0_higher_energy}
    \end{subfigure}
    \hfill
    \begin{subfigure}[b]{0.24\linewidth}
    \centering
    \includegraphics[width=\linewidth]{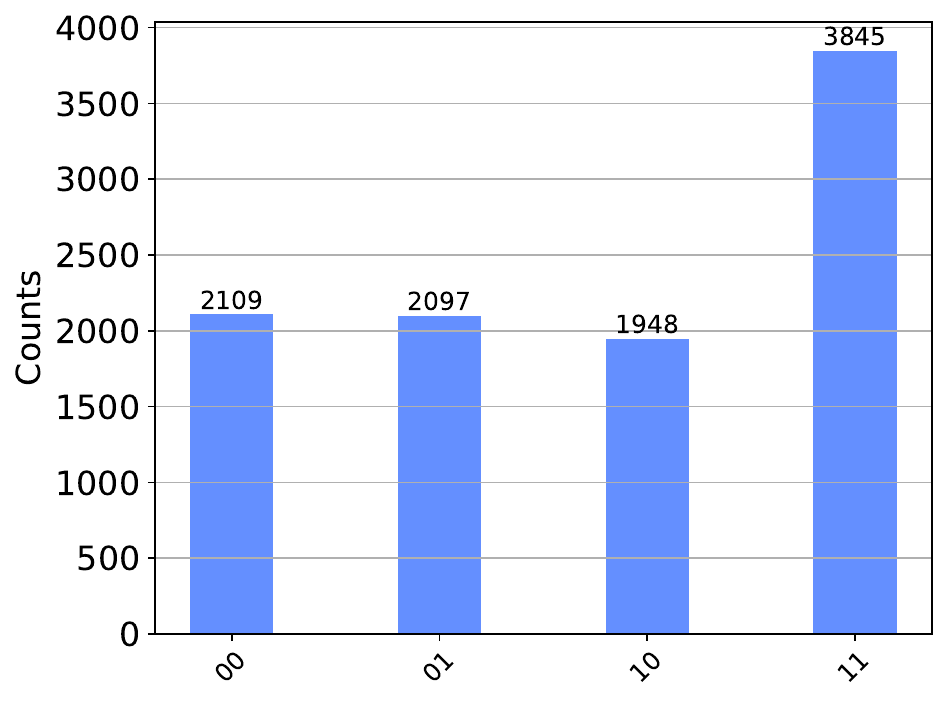}
    \caption{HEA on HHL clock1 using qubit 0,1,2,3 on \textit{IBM\_brisbane}.}
    \label{subfig:attack_clock1_higher_energy}
    \end{subfigure}
    \hfill
    \begin{subfigure}[b]{0.24\linewidth}
    \centering
    \includegraphics[width=\linewidth]{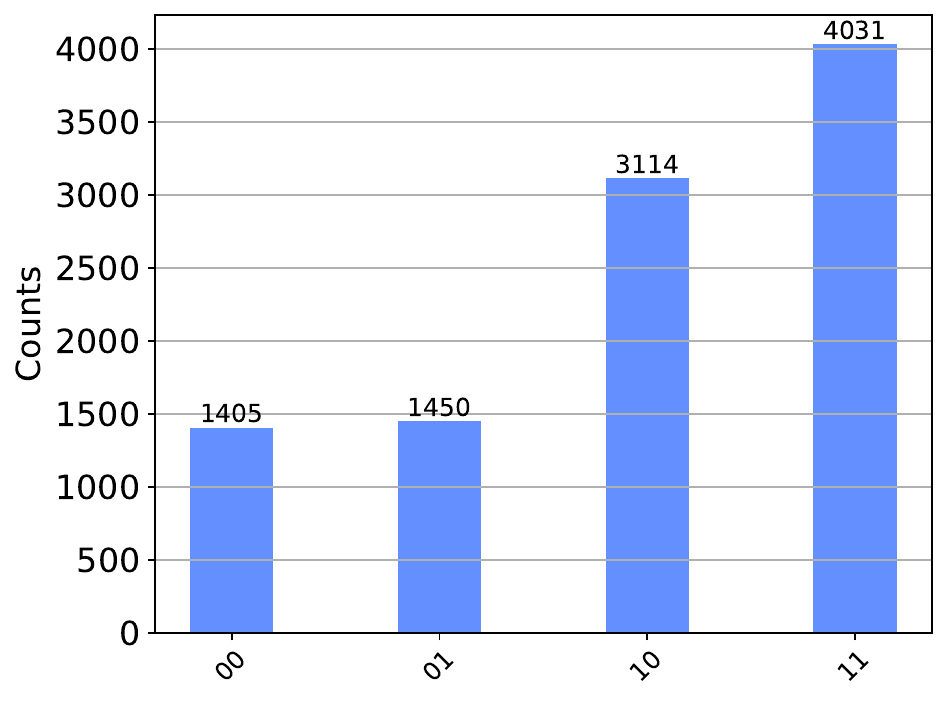}
    \caption{HEA on HHL b using qubit 0,1,2,14 on \textit{IBM\_brisbane}.}
    \label{subfig:attack_b_higher_energy}
    \end{subfigure}
    \vskip\baselineskip
    \begin{subfigure}[b]{0.24\linewidth}
    \centering
    \includegraphics[width=\linewidth]{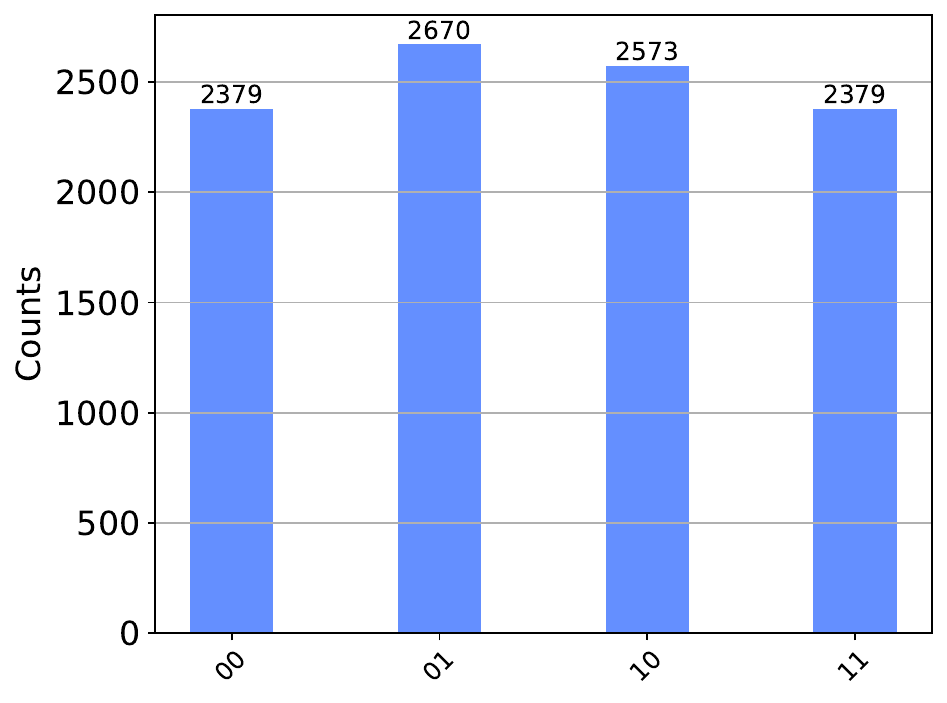}
     \caption{IIA on HHL ancilla using qubit 0,1,2,3 on \textit{IBM\_brisbane}.}
     \label{subfig:attack_a_a=1}
    \end{subfigure}
    \begin{subfigure}[b]{0.24\linewidth}
    \centering
    \includegraphics[width=\linewidth]{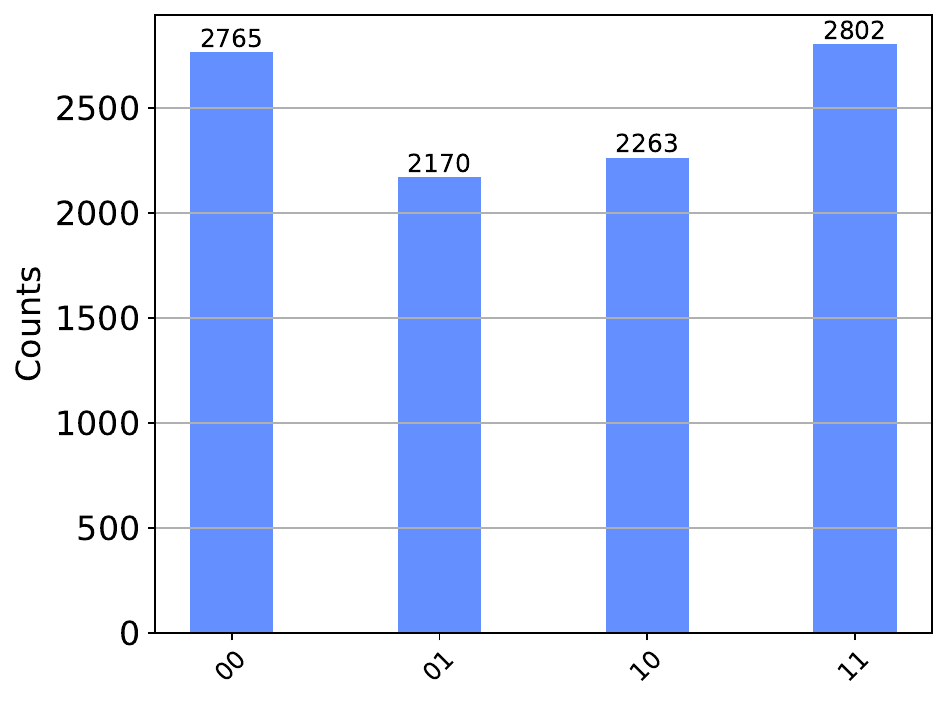}
     \caption{IIA on HHL clock0 using qubit 0,1,2,3 on \textit{IBM\_brisbane}.}
     \label{subfig:attack_clock0_clock0=1}
    \end{subfigure}
    \begin{subfigure}[b]{0.24\linewidth}
    \centering
    \includegraphics[width=\linewidth]{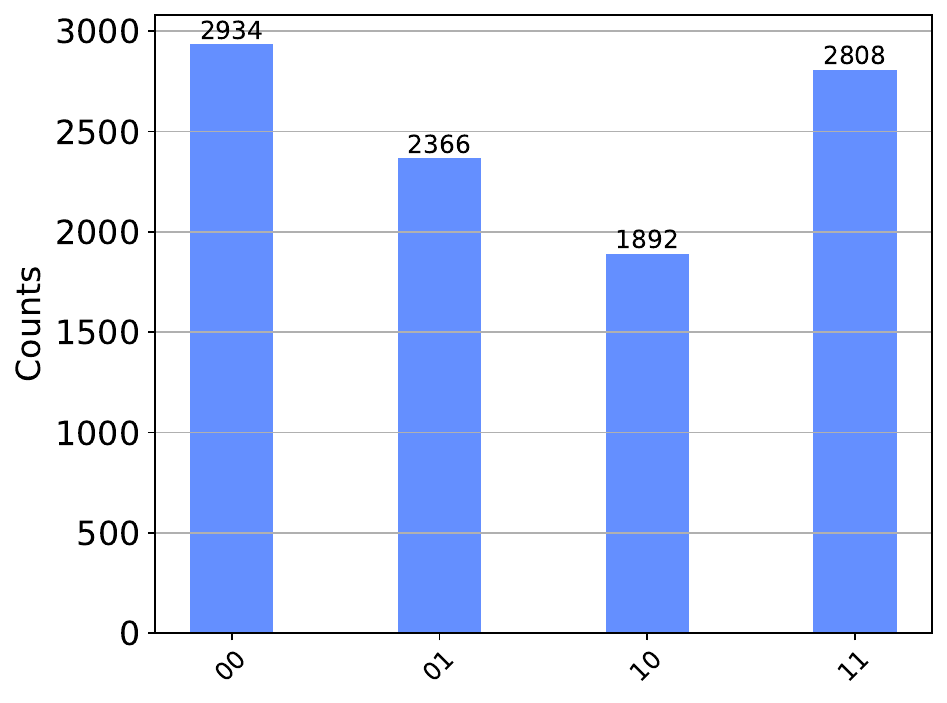}
     \caption{IIA on HHL clock1 using qubit 0,1,2,3 on \textit{IBM\_brisbane}.}
     \label{subfig:attack_clock1_clock1=1}
    \end{subfigure}
    \begin{subfigure}[b]{0.24\linewidth}
    \centering
    \includegraphics[width=\linewidth]{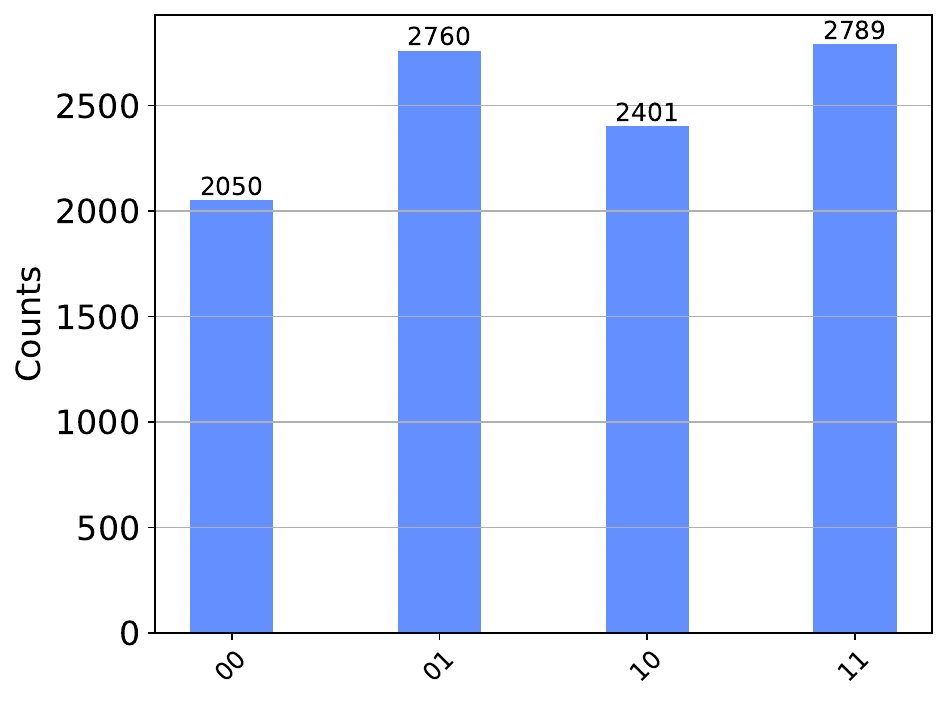}
     \caption{IIA on HHL b using qubit 0,1,2,14 on \textit{IBM\_brisbane}.}
     \label{subfig:attack_b_b=1}
    \end{subfigure}
    \caption{HEA and IIA on different HHL qubits on \textit{IBM\_brisbane}. The graph represents the average output of three tests.}
    \label{fig:enter-label}
\end{figure*} 

\begin{figure}[t]
 \centering
 \includegraphics[width=0.24\textwidth]{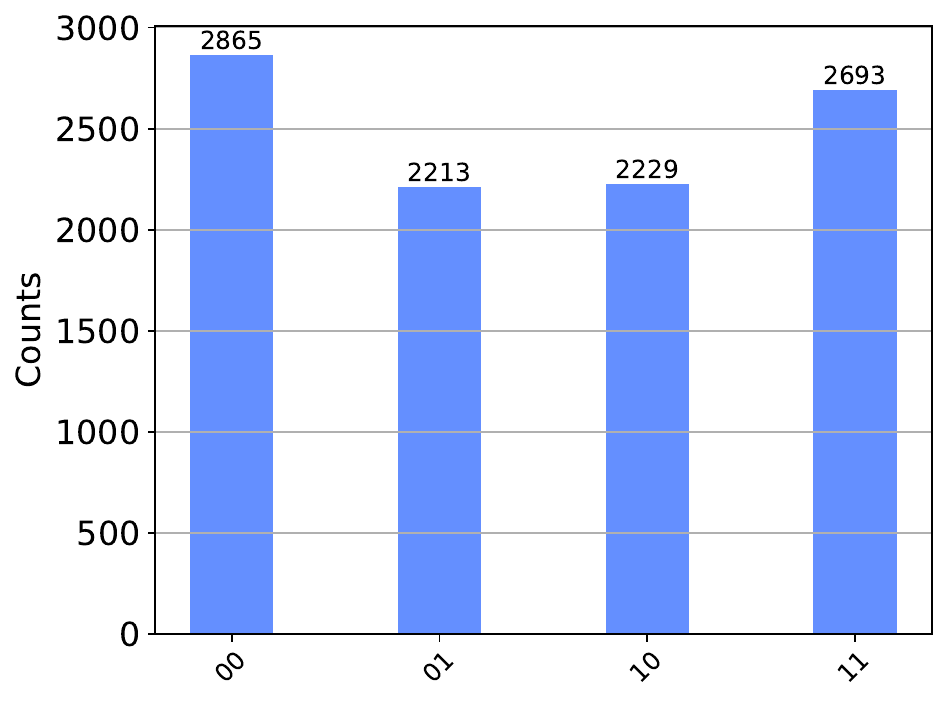}
 \caption{HEA on HHL all 4 qubits using qubit 0,1,3,14 on \textit{IBM\_brisbane}.}
 \label{fig:hea_4_qubits}
\end{figure}

\begin{figure}[t]
 \centering
 \includegraphics[width=0.24\textwidth]{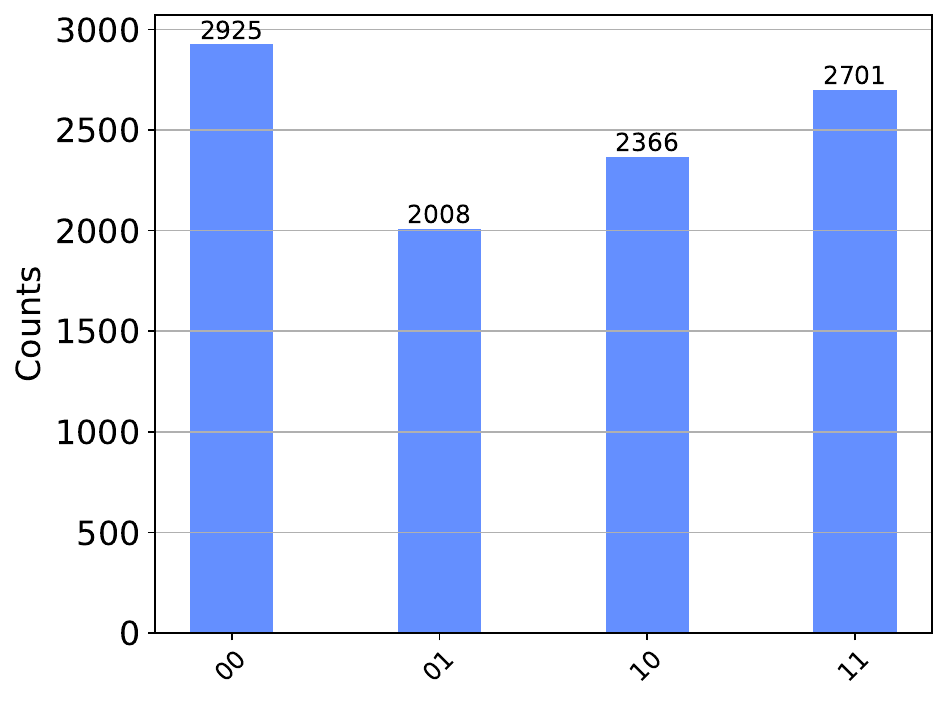}
 \caption{IIA on HHL all 4 qubits using qubit 0,1,3,14 on \textit{IBM\_brisbane}.}
 \label{fig:iia_4_qubits}
\end{figure}

\subsubsection{IIA and HEA on Ancilla Qubit}

Fig.~\ref{subfig:attack_a_higher_energy} shows the HEA on HHL ancilla qubit. This attack greatly harms the algorithm output and reduces the ratio to 1:0.8095. IIA is meanwhile presented in Fig.~\ref{subfig:attack_a_a=1} and it also decreases the ratio to 1:0.8910.

\subsubsection{IIA and HEA on Clock Qubits}

Fig.~\ref{subfig:attack_clock0_higher_energy} and Fig.~\ref{subfig:attack_clock0_clock0=1} show the two attacks on HHL clock0 qubit. The HEA on clock0 decreases the ratio to 1:0.6849, while IIA on clock0 does not affect HHL that much, with ratio of 1:1.2912.
Attacks on HHL clock1 qubit in Fig.~\ref{subfig:attack_clock1_higher_energy} and Fig.~\ref{subfig:attack_clock1_clock1=1} show interesting influence on the HHL output. The ratio of HEA on clock1 is 1:1.1835. The ratio of IIA on clock1 is 1:1.1868, which is almost the same as original HHL output.

\subsubsection{IIA and HEA on b Qubit}

Fig.~\ref{subfig:attack_b_higher_energy} present the HEA on HHL b qubit. This attack is allocated to qubit 0, 1, 2, 14 on $IBM\_brisbane$ machine. The ratio for the HEA on b is 1:2.7800. Even though it seems that the HEA on b improves HHL performance closer to classical solution, this attack will actually force the ratio higher than classical solution if we have better quantum hardwares than NISQ devices. The ratio for IIA on HHL b qubit in Fig.~\ref{subfig:attack_b_b=1} is 1:1.0105.

\subsubsection{IIA and HEA on all Qubits}

Fig.~\ref{fig:hea_4_qubits} and Fig.~\ref{fig:iia_4_qubits} show the two attacks on all HHL qubits. The attacks do not affect HHL ratios that much, while they have impact on the output distribution, with HEA variational distance of 0.1191 and IIA variational distance of 0.1388

\subsection{Summary of the Effectiveness of the Attacks}

Table~\ref{tab:higher_energy_attack_hardware} and table~\ref{tab:improper_initialization_attack_hardware} summarize these two attacks on HHL algorithm on $IBM\_brisbane$ machine. We use the average ratio and the variational distance to quantify the difference between probability distribution of original HHL hardware output and that of attacked output. On real hardware, HEA seem to have bigger impact in terms of the variational distance metric.
In table~\ref{tab:hea_iia_attack}, we evaluate the success of an attack by determining whether it causes the ratio to greatly deviate from the ratio observed without an attack. As shown in the table, all the HEA and IIA succeeded except for IIA on clock1, demonstrating the overall effectiveness of our HEA and IIA.
On the other hand, the noisy nature of the NISQ computers means that effects of the attacks and the noise both affect the output, and further study of the attacks on real hardware is necessary.

%% file: sections/hardware_defense.tex
\section{Evaluation of the Defense}

\begin{table*}[t]
    \centering
    \caption{\small Summary of HHL outputs without defense under HEA and IIA on $IBM\_brisbane$. We test attacks on the original HHL circuit. The checkmark represents the success of the attack detection while the cross means it fails. For the defense, we assume that the defense is successful if the ratio with defense is closer to the baseline (without attack) of 1:1.1759 or the theoretical ratio of 1:9, compared to the outcome without~defense.}
    \small
    \begin{tabular}{c|c|c|c|c|c}
    \hline
    \multicolumn{2}{c|}{\textbf{Victim Qubits}} & \multicolumn{2}{c|}{\textbf{HEA}} & \multicolumn{2}{c}{\textbf{IIA}}\\ \hline
    \textbf{Num. Victim Qubits} & \textbf{Victim} & \textbf{Ratio} & \textbf{Attack Succeeded} & \textbf{Ratio} & \textbf{Attack Succeeded}\\ \hline \hline
    0 & \textbf{No attack} & 1 : 1.1883 & \textbf{-} & 1 : 1.1759 & \textbf{-} \\ \hline  \hline
    1 & \textbf{ancilla} & 1 : 0.8095 & \textbf{\Checkmark} & 1 : 0.8910 & \textbf{\Checkmark} \\ \hline
    1 & \textbf{clock0} & 1 : 0.6849 & \textbf{\Checkmark} & 1 : 1.2912 & \textbf{\Checkmark} \\ \hline
    1 & \textbf{clock1} & 1 : 1.8335 & \textbf{\Checkmark} & 1 : 1.1868 & \textbf{\XSolidBrush} \\ \hline
    1 & \textbf{b} & 1 : 2.7800 & \textbf{\Checkmark} & 1 : 1.0105 & \textbf{\Checkmark} \\ \hline \hline
    4 & \textbf{All HHL qubits} & 1 : 2.2500 & \textbf{\Checkmark} & 1 : 1.2169 & \textbf{\Checkmark} \\ \hline
    \end{tabular}
    %}
    \label{tab:hea_iia_attack}
\end{table*}

\begin{table*}[t]
    \centering
    \caption{\small Summary of HHL with defense outputs under HEA and IIA on $IBM\_brisbane$. We test attacks on the defense circuit and compare the output we get with the expected output. The 7-bit output of the detection registers includes, from left to right, 2 bits for {\tt c\_ancilla\_defense}, 3 bits for {\tt c\_b\_defense}, and 2 bits for {\tt c\_clock\_defense}. The baseline expected outputs are {\tt 10 000 00} when HHL converges and {\tt 01 000 00} when HHL is still updating. The checkmark represents the success of the attack detection while the cross means it fails.}
    \small
    \begin{tabular}{c|c|c|c|c|c}
    \hline
    \multicolumn{2}{c|}{\textbf{Victim Qubits}} & \multicolumn{2}{c|}{\textbf{HEA}} & \multicolumn{2}{c}{\textbf{IIA}}\\ \hline
    \textbf{Num. Victim Qubits} & \textbf{Victim} & \textbf{Actual Output} & \textbf{Defense Succeeded} & \textbf{Output of Detection Registers} & \textbf{Defense Succeeded}\\ \hline
    0 & \textbf{No attack} & {\tt 10 000 00} & \textbf{\Checkmark} & {\tt 10 000 00} & \textbf{\Checkmark} \\ \hline \hline
    1 & \textbf{ancilla} & {\tt 11 010 00} & \textbf{\Checkmark} & {\tt 01 001 00} &\textbf{\Checkmark} \\ \hline
    1 & \textbf{new ancilla} & {\tt 01 111 00} & \textbf{\Checkmark} & {\tt 01 100 00} &\textbf{\Checkmark} \\ \hline
    1 & \textbf{clock0} & {\tt 11 011 11} & \textbf{\Checkmark} & {\tt 01 001 01} &\textbf{\Checkmark} \\ \hline
    1 & \textbf{clock1} & {\tt 11 011 11} & \textbf{\Checkmark} & {\tt 10 001 00} &\textbf{\Checkmark} \\ \hline
    1 & \textbf{b} & {\tt 01 001 11} & \textbf{\Checkmark} & {\tt 01 001 00} &\textbf{\Checkmark} \\ \hline \hline
    4 & \textbf{All HHL qubits} & {\tt 10 011 01} & \textbf{\Checkmark} & {\tt 10 010 10} &\textbf{\Checkmark} \\ \hline
    \end{tabular}
    \label{tab:hea_iia_defense}
\end{table*}

Table~\ref{tab:hea_iia_defense} compares the performance of our defense circuit under attack against its baseline functionality.
The first two columns summarize the victim qubits being attacked. Columns 3 and 5 denote the actual outputs obtained by our experiments for HEA and IIA with and without attack, respectively, while Columns 4 and 6 indicate whether the attacks were successful.
The 7-bit output of the detection registers includes all the measurements from the defense mechanisms  as shown in part 2, 7, and 11 in Fig.~\ref{fig_defense_circuit}, which are, from left to right, 2 bits for {\tt c\_ancilla\_defense}, 3 bits for {\tt c\_b\_defense}, and 2 bits for {\tt c\_clock\_defense}. The measurement of the b qubit in part 10 of Fig.\ref{fig_defense_circuit} is excluded, as our focus is solely on testing the defense circuit rather than the actual ratio of the HHL algorithm.

The baseline outputs are {\tt 10 000 00} when HHL converges and {\tt 01 000 00} when HHL is still iterating, all other outputs indicate some sort of error.
For our defense mechanism, we consider the defense is successful if the actual output deviates from the baseline output, indicating the detection of an attack.  As shown in Table~\ref{tab:hea_iia_defense}, the actual outputs match the baseline when no attack is present, demonstrating that our mechanism does not interfere with the proper functioning of the HHL algorithm. This confirms that our defense strategy, which employs an additional qubit, maintains a high level of output accuracy. Furthermore, Table~\ref{tab:hea_iia_defense} shows that our defense successfully detects all types of HEA and IIA targeting various victim qubits. Even attacks on multiple qubits are effectively mitigated by our defense mechanism as shown in the last row of the tables where all $4$ HHL qubits are attacked.

Our defense mechanism can further distinguish, to some extent, which victim qubits the attacker is targeting by combining the results from table~\ref{tab:defense_ancilla},~\ref{tab:defense_b} and~\ref{tab:defense_clock}. For instance, outputs such as {\tt 01 001 00} or {\tt 10 001 00} indicate IIA on the ancilla qubit, while {\tt 01 100 00} or {\tt 10 100 00} signify IIA on the new ancilla qubit. These outputs align with the experimental results shown in Table~\ref{tab:hea_iia_defense}.
However, not all actual outputs match the expected outputs predicted by table~\ref{tab:defense_ancilla},~\ref{tab:defense_b} and~\ref{tab:defense_clock}. Possible reasons for these discrepancies include noise and the unique characteristics of higher-energy states.

Given the characteristics of higher energy states, it is essential to ensure that after transpilation, the attack pulse and measurement of the corresponding victim qubit are mapped to the same physical qubit on $IBM\_brisbane$. In our experiments, we addressed this issue by prioritizing circuit correctness over optimization levels. This involved meticulous inspection of each transpiled circuit and forgoing some optimization steps. While this approach can be further improved by selectively choosing qubits with optimal connectivity for HHL, ensuring that all 7-bit measurement outputs map to the same physical qubits remains difficult on $IBM_brisbane$ due to the limitations of its transpilation process.
Second, due to our limited access to the IBM machine, we were only able to conduct frequency sweep and Rabi experiments on qubits 0, 1, 2, 3, and 14 to construct the HEA pulse. This constraint restricted our experiments to these five qubits. Future work could benefit from exploring additional qubits with lower error rates to improve the outputs of the HHL~circuit.

\subsection{Resilience of Defense to Noise}

\begin{figure}[t]
 \centering
 \includegraphics[width=0.46\textwidth]{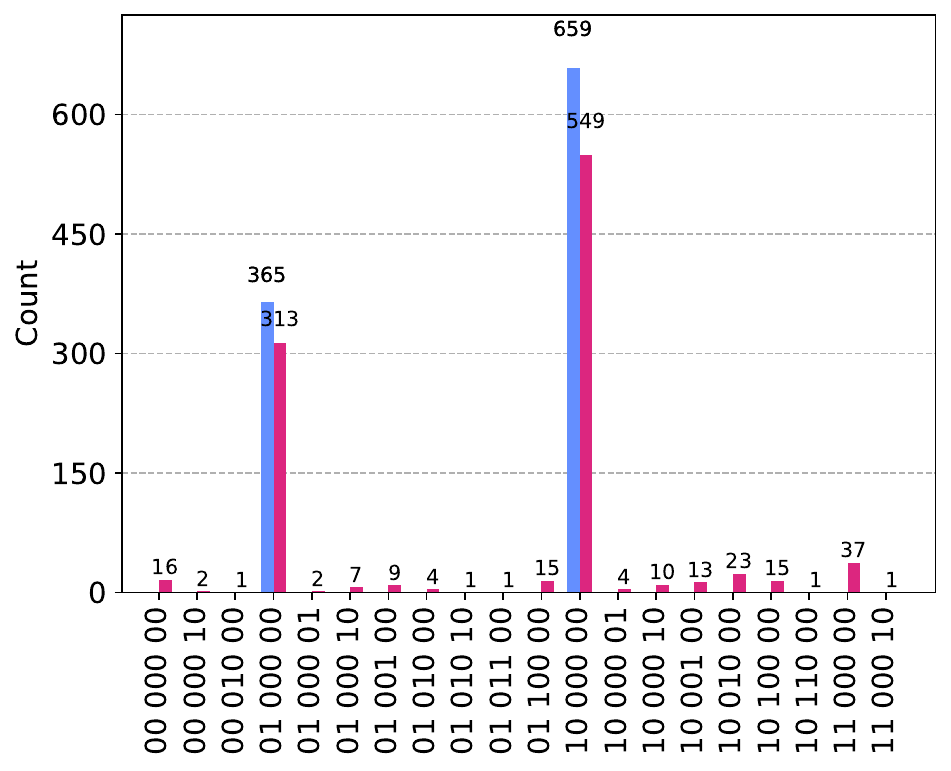}
 \caption{HHL defense experiments using qubit 0,1,2,3,14 on ideal simulator and fake backend $fake\_brisbane$ with 1024 shots. The baseline expected outputs are {\tt 10 000 00} when HHL converges and {\tt 01 000 00} when HHL is still iterating through the optimization process. All other outputs represent noisy~outputs.}
 \label{fig_defense_ideal_and_fake_backend}
\end{figure}

We have further evaluated our defense mechanism to ensure that it works correctly in presence of the noise in today's Noisy Intermediate Scale Quantum (NISQ) computers. Fig.~\ref{fig_defense_ideal_and_fake_backend} shows the output of the HHL algorithm with our defense when there is no attack. The baseline is the ideal simulator shown as blue bars. As can be seen the dominant outputs of the attack detection registers are {\tt 10 000 00} when HHL converges and {\tt 01 000 00} when HHL is still iterating, while no incorrect outputs are present. The red bars show the execution on simulation simulating the $IBM\_brisbane$. Simulation of the $IBM\_brisbane$ is typically called $FakeBrisbane$ by IBM. When the simulation includes noise emulating the execution of the $IBM\_brisbane$, we observe a limited amount of incorrect (noisy) output. They are limited and can be discarded, and the correct {\tt 10 000 00} and {\tt 01 000 00} outputs remain dominant. Thus our defense is resilient to noise, which should be expected as we do not introduce significant new gates into the circuit nor do we increase the depth of the HHL circuit by any noticeable amount.

%% file: sections/upgraded_hhl.tex
\subsection{Application to Other HHL Versions}

The HHL algorithm was created as a quantum counterpart to classical systems of linear equation solvers. Since solving large systems of linear equations represents a significant bottleneck not only in machine learning but also in engineering, the intention behind the original HHL algorithm was to provide an exponential speedup over the best classical algorithms. 
Recently, the HHL algorithm has been further improved in terms of its time complexity. Ambainis~\cite{ambainis2010variabletimeamplitudeamplification} was able to achieve a complexity that is linear in 
$k$, resulting in a time complexity of $O(log(N) s^2 k/\epsilon)$.
Furthermore, the work of Childs et al.~\cite{Childs_2017} improved the HHL algorithm even further, making its time complexity $O(sk log(N) poly log(1/\epsilon))$.
Finally, Wossnig et al.~\cite{Wossnig_2018} improved the original HHL algorithm, achieving a time complexity of $O(poly log (N) ||A||_{F} k^2/\epsilon)$.
There have also been improvements in terms of circuit depth and efficiency regarding the original HHL algorithm. Cao et al.~\cite{Cao_2012} proposed an efficient quantum circuit implementation using only elementary quantum operations. Lee et al.~\cite{Lee_2019} designed a hybrid HHL algorithm, demonstrating that their hybrid approach can reduce the number of two-qubit gates in the circuit. Furthermore, Zhang et al.~\cite{Zhang2022} combined the works of Cao et al. and Lee et al., demonstrating higher fidelity in the experimental solution along with reduced quantum resource consumption. Moreover, Yalovetzky et al.~\cite{yalovetzky2024solvinglinearsystemsquantum} extended the work of Lee et al. by optimizing the scaling parameter $\gamma$ for Hamiltonian simulation. They also introduced a new variant of quantum phase estimation called semiclassical quantum phase estimation. Finally, Perelshtein et al.~\cite{https://doi.org/10.1002/andp.202200082} designed and implemented a quantum hybrid HHL algorithm focusing on phase estimation and the classical optimization of circuit width and depth.

We believe our generic defense ideas presented in Sec.~\ref{sec_defense_idea} can be applied to these variants of HHL as the defenses are specific to IIA and HEA, and not to the algorithm being protected. Currently, however, to the best of our knowledge, the other HHL variants do not have publicly available code and we could not prototype defnese for these HHL variants. This is left as future work.

%% file: sections/related_work.tex
\section{Related Work}

The security of quantum algorithms and quantum information, in general, has become a very interesting field lately. Given that quantum machines are developing at a rapid pace, researchers have started focusing on the security aspects. Therefore,~\cite{Yang:2023nec} exhibits how to securely transmit information using the HHL algorithm to prevent information leakage, while~\cite{Farokhi:2023zei} defines a new measure of information leakage for the quantum encoding of classical data. Error propagation in the HHL algorithm has been studied in~\cite{9908231}, where the authors identified three major sources of errors: single-qubit flipping, gate infidelity, and error propagation.
There is also a potential way to reduce the demands on physical qubits by evaluating the resource cost of quantum phase estimation, which is a crucial part of the HHL algorithm, before and after quantum error correction~\cite{Zheng:2024pyx}. Similarly, quantum phase estimation, being one of the most computationally expensive components of the HHL algorithm, has been tested in terms of scaling properties and related noise resilience~\cite{marfany2024identifyingbottlenecksnisqfriendlyhhl}.

On the security attack side, researchers are actively exploring software supply chain and other attacks on quantum computers. Prior work has demonstrated that the gate-level to pulse-level specification of circuits could be abused to inject attacks in quantum circuits~\cite{2406.05941}. This is an example of software supply chain attacks that could be leveraged to deploy the IIA and cause HHL to generate incorrect results. In parallel, researchers have explored higher energy state attacks~\cite{mi2022securing}. The HEA used in our work directly leverages higher energy state attacks~\cite{mi2022securing}, but applies them to a new victim quantum circuit algorithm, the HHL.

%% file: sections/conclusion_future_work.tex
\section{Conclusion and Future Work}

This work demonstrated two types of attacks that could be performed on the HHL algorithm, both on quantum simulators and on quantum hardware. To address the attacks, this work presented novel defense strategies against these attacks. To the best of our knowledge, this is the first work that shows attacks and explains possible defense strategies for the HHL algorithm, both in theory and in practice. It further demonstrates a practical defense circuit that detects the attack with limited overhead and is resilient to~noise.

For future research, there are several directions that can be pursued thanks to the new understanding of attacks and defenses presented in this work. One new direction is investigating whether more efficient alternatives, in terms of circuit design, exist that could achieve the same goals with fewer resources. We believe our design to be minimal, but further optimization may be possible. Another path would involve exploring newer HHL algorithm versions (both fully quantum and hybrid) and applying existing defense strategies on~them. We have not been able to access code for any newer HHL algorithm versions, but our work may motivate researchers working on HHL to share their code and enable further~evaluation.

%% file: sections/acknowledgements.tex
    \section*{Acknowledgments}

This work was supported in part by NSF grant \nsf{2332406}.